\documentclass[%
 reprint,
 amsmath,amssymb,
 aps,
]{revtex4-2}

\usepackage{color}
\usepackage{graphicx}% Include figure files
\usepackage{float}
\usepackage{dcolumn}% Align table columns on decimal point
\usepackage{tabularx}
\usepackage{bm}% bold math
\usepackage{hyperref}% add hypertext capabilities
\usepackage{soul}
\usepackage{orcidlink}
\usepackage{amsmath}
\usepackage{amssymb}
\usepackage{cancel}
\usepackage{array}     % For >{\command} column specifiers if needed
\usepackage{makecell}  % For manual line breaks within cells and better control

\hypersetup{colorlinks = true,
            linkcolor = blue,
            anchorcolor = blue,
            citecolor = blue,
            filecolor = blue,
            urlcolor = blue
            }
\usepackage{tikz,xcolor,hyperref}% Make Orcid icon
\definecolor{lime}{HTML}{A6CE39}
\DeclareRobustCommand{\orcidicon}{%
    \begin{tikzpicture}
    \draw[lime, fill=lime] (0,0) 
    circle [radius=0.16] 
    node[white] {{\fontfamily{qag}\selectfont \tiny ID}};    \draw[white, fill=white] (-0.0625,0.095) 
    circle [radius=0.007];    \end{tikzpicture}
    \hspace{-2mm}}
    \foreach \x in {A, ..., Z}{%
    \expandafter\xdef\csname orcid\x\endcsname{\noexpand\href{https://orcid.org/\csname orcidauthor\x\endcsname}{\noexpand\orcidicon}}
}

\begin{document}

\preprint{APS/123-QED}

\title{Multiband parameter estimation with phase coherence and extrinsic marginalization: \\Extracting more information from low-SNR CBC signals in LISA data}

\author{Shichao Wu\orcidlink{0000-0002-9188-5435}$^{1}$}
\email{shichao.wu@aei.mpg.de} % Footnote email for Author 1
\author{Alexander H. Nitz\orcidlink{0000-0002-1850-4587}$^{2}$}
\email{ahnitz@syr.edu}   % Footnote email for Author 2
\author{Ian Harry\orcidlink{0000-0002-5304-9372}$^{3}$}
\author{Stanislav Babak\orcidlink{0000-0001-7469-4250}$^{4}$}
\author{Michael J. Williams\orcidlink{0000-0003-2198-2974}$^{3}$}
\author{Collin Capano\orcidlink{0000-0002-0355-5998}$^{2,5}$}
\author{Connor Weaving\orcidlink{0009-0008-2697-2998}$^{3}$}
    % Using orcidlink package convention here

\affiliation{ % Start of single affiliation block
$^1$ Max-Planck-Institut f{\"u}r Gravitationsphysik (Albert-Einstein-Institut) and Leibniz Universit{\"a}t Hannover, D-30167 Hannover, Germany \\
$^2$ Department of Physics, Syracuse University, Syracuse NY 13244, USA \\
$^3$ University of Portsmouth, Institute of Cosmology and Gravitation, Portsmouth PO1 3FX, United Kingdom \\
$^4$ Astroparticule et Cosmologie, Université de Paris, CNRS, 75013 Paris, France \\
$^5$ Department of Physics, University of Massachusetts Dartmouth, North Dartmouth, MA 02747, USA
} % End of single affiliation block

\date{\today}

\begin{abstract}

This paper presents a novel coherent multiband analysis framework for characterizing stellar- and intermediate-mass binary black holes using LISA and next-generation ground-based detectors (ET and CE), leveraging the latest developments in the \texttt{PyCBC} pipeline. Given the population parameters inferred from LVK results and LISA's sensitivity limits at high frequencies, most stellar-mass binary black holes would likely have SNRs below 5 in LISA, but the most state-of-the-art multiband parameter estimation methods, such as those using ET and CE posteriors as priors for LISA, typically struggle to analyze sources with a LISA SNR less than 5. We present a novel coherent multiband parameter estimation method that directly calculates a joint likelihood, which is highly efficient; this efficiency is enabled by multiband marginalization of the extrinsic parameter space, implemented using importance sampling, which can work robustly even when the LISA SNR is as low as 3. Having an SNR of $\sim 3$ allows LISA to contribute nearly double the number of multiband sources. Even if LISA only observes for one year, most of the multiband detector-frame chirp mass's 90\% credible interval (less than $10^{-4} \mathrm{M}_\odot$) is still better than that of the most accurately measured events for ET+2CE network in 7.5 years of observation, by at least one order of magnitude. For the first time, we show efficient multiband Bayesian parameter estimation results on the population scale, which paves the way for large-scale astrophysical tests using multibanding. 

\end{abstract}

\maketitle

\section{\label{sec:intro}Introduction\protect\\}

The Laser Interferometer Space Antenna (LISA) is anticipated to capture gravitational wave (GW) signals from stellar-mass binary black holes (sBBHs). Current ground-based observatories confirm an extragalactic population of black hole binaries, with component masses from $5M_\odot$ to at least $85 M_\odot$ \cite{KAGRA:2021vkt, Nitz:2021zwj, Olsen:2022pin}. Over 100 such systems, corresponding to their final inspiral, merger, and ringdown, have been confidently identified in the first three observing runs (O4 itself might double this). These signals last from hundreds of milliseconds to seconds in the LIGO-Virgo-KAGRA (LVK) band, while their early inspiral would persist in LISA's band for over its nominal 4-year mission \cite{Seoane:2021kkk}. However, detecting these faint signals in instrumental noise is a major challenge \cite{Moore:2019pke}. The scarcity of observed high-mass black hole binaries creates significant uncertainty in estimating merger rates for massive systems \cite{KAGRA:2021duu, Nitz:2021zwj, Roulet:2021hcu}. This uncertainty propagates into predictions for the LISA-alone detectable population, with models projecting 0 to 3.6 massive events ($m_{1}>50 M_\odot$) during its operation with LISA signal-to-noise ratios (SNR) greater than 8 (see Table 3.8 in \cite{LISA:2024hlh}). Some sBBH systems may be observed across multiple frequency bands: early inspiral by LISA, then later by ground-based detectors after leaving LISA's sensitivity band \cite{Sesana:2016ljz}, over months to years.

Under favorable conditions, early LISA alerts for sBBHs \cite{Sesana:2016ljz} could help localize and predict merger events for ground-based follow-up, potentially improving prospects for identifying their electromagnetic counterparts \cite{Perna:2016jqh, Mink:2017npg}. While detecting such systems with LISA alone is estimated to be very limited or impossible \cite{Moore:2019pke}, multiband observations using ground-based detector information would improve precision in determining intrinsic (e.g., mass, spin) and extrinsic (e.g., sky location, distance) parameters \cite{Toubiana:2022vpp}. However, population models predict just one or two such multiband sources from the nominal 4-year LISA mission with a conservative LISA SNR threshold of 8 and reasonable waiting time \cite{Toubiana:2022vpp, Seto:2022xmh, LISA:2024hlh, Buscicchio:2024asl}. In contrast, GW190521-like or more massive intermediate-mass black hole binaries (IMBHB) could be promising for multiband studies \cite{Jani:2019ffg}, if their merger rates are high enough. Although their higher LISA SNR and shorter merger timescales enhance detectability with next-generation ground-based detectors, feasibility depends on their actual event rates.

Multiple formation channels are proposed for sBBHs \cite{Mandel:2021smh}. LISA can uniquely distinguish imprints of different formation mechanisms by analyzing orbital eccentricity and component spin orientations \cite{Nishizawa:2016jji, Nishizawa:2016eza, Breivik:2016ddj, Kremer:2018cir, Samsing:2018isx, DOrazio:2018jnv, Wang:2023tle}. This is because GW-driven orbital circularization \cite{Peters:1964zz} makes eccentricity measurements more accessible in LISA's frequency regime, unlike ground-based detectors observing at higher frequencies where eccentricity is already significantly reduced \cite{Nishizawa:2016eza}. Precise timing from third-generation (3G) ground-based detectors can resolve degeneracies between coalescence time and eccentricity from LISA data \cite{Klein:2022rbf}. Precise multiband measurement of the remnant's recoil kick can reveal their formation channel and enable multimessenger follow-up for events in active galactic nucleus disks \citep{Ranjan:2024wui}.

sBBHs are critical for source population studies and fundamental physics. Multiband observations improve cosmological constraints via the dark siren method \cite{Muttoni:2021veo,Seymour:2022teq}, can constrain GW speed to $\sim 10^{-17}$ (deviations from GR indicating new physics \cite{Baker:2022eiz, Harry:2022zey}), and offer unique constraints on dipole radiation and graviton mass \cite{Barausse:2016eii, Toubiana:2020vtf}. Combining multiband parameter estimation with parameterized post-Newtonian (ppN) formalisms enables sub-percent precision constraints on post-Newtonian phasing coefficients \cite{Gupta:2020lxa}; similarly, the parametrized post-Einsteinian (ppE) approach significantly benefits from multiband observations, enhancing tests of gravitational theories \cite{Barausse:2016eii, Carson:2019kkh, Gnocchi:2019jzp}. For individual systems, joint analysis of LISA’s early inspiral and ground-based late-inspiral/merger/ringdown observations allows rigorous inspiral-merger-ringdown (IMR) consistency tests \cite{Carson:2019kkh}. Optimally, LISA early warnings could optimize ground-based detector readiness for ringdown signals or ensure observatory availability for critical mergers \cite{Tso:2018pdv}. Gravitational memory effects from sBBHs may be detectable in LISA's band using triggers from future ground-based interferometers \cite{Ghosh:2023rbe}.

Robust data analysis methodologies are required to achieve these scientific objectives. Blind matched-filtering searches require prohibitively large computational resources (some studies suggest $\mathcal{O}(10^{41})$ templates \cite{Moore:2019pke}). A viable solution uses ground-based detector constraints to narrow the parameter space, enabling archival analysis that minimizes computational demands, reduces the LISA SNR detection threshold, and increases detection rates $\sim 4-8$ times \cite{Wong:2018uwb, Gerosa:2019dbe, Cutler:2019krq, Ewing:2020brd} compared to blind searches (LISA SNR threshold 8). Semi-coherent searches with heuristic parameter optimization \citep{Bandopadhyay:2023gkb,Bandopadhyay:2024lwv,Bandopadhyay:2024nys} can use LISA data alone but only reach LISA SNR down to $\sim 12$ for sBBHs. Most multiband parameter estimation studies used the Fisher information matrix (FIM), unsuitable for low SNR LISA signals \cite{Vallisneri:2007ev,Liu:2020nwz}. Bayesian inference is critical to accurately estimate sBBH properties \cite{Toubiana:2020cqv, Marsat:2020rtl, Buscicchio:2021dph, Digman:2022igm}. However, low SNR LISA-observed sBBHs \cite{Moore:2019pke} and complex likelihood surfaces challenge traditional Bayesian parameter estimation. To infer source parameters, the 3G posterior can be a prior for LISA Bayesian estimation \citep{Toubiana:2022vpp}, but this struggles with sBBH signals less than LISA SNR 5, as samplers can easily miss the true likelihood among numerous local maxima.

This paper introduces a novel coherent multiband method to overcome these difficulties, analyzing the joint LISA and ground-based detector posterior simultaneously and efficiently. For the first time, (1) we extract useful information from sBBHs in LISA's band down to LISA SNR $\sim 3$, almost doubling the multiband event rate; (2) we perform multiband Bayesian parameter estimation on the population scale. Key improvements include using rotation matrices to unify LISA and 3G signal parameter spaces and importance sampling to marginalize extrinsic parameters, reducing parameter space dimensionality and complexity. We also extend the heterodyned technique \citep{Cornish:2010kf, Zackay:2018qdy, Finstad:2020sok, Cornish:2021lje} to accelerate the joint multiband likelihood.

This paper is structured as follows. Sec.~\ref{sec:background} presents a potential LISA+3G multiband observation scenario: detailing selected detectors, sensitivities, multiband event rates, and sBBH signal time evolution across LISA and 3G bands. Sec.~\ref{sec:method} introduces our novel coherent multiband Bayesian parameter estimation method. Sec.~\ref{sec:population_runs} further validates our method on multiband sBBH and IMBHB systems from the \texttt{GWTC-3} population model; for IMBHB, we also check if higher modes improve results. We perform population-scale multiband Bayesian parameter estimation down to LISA SNR $\sim 3$ for the first time. Finally, Sec.~\ref{sec:conclusions} summarizes our methodology and results, discusses potential improvements, and highlights future studies benefiting from this framework.

\begin{figure*}
    \centering
    \vspace*{-0.5cm}
    \includegraphics[width=1\textwidth]{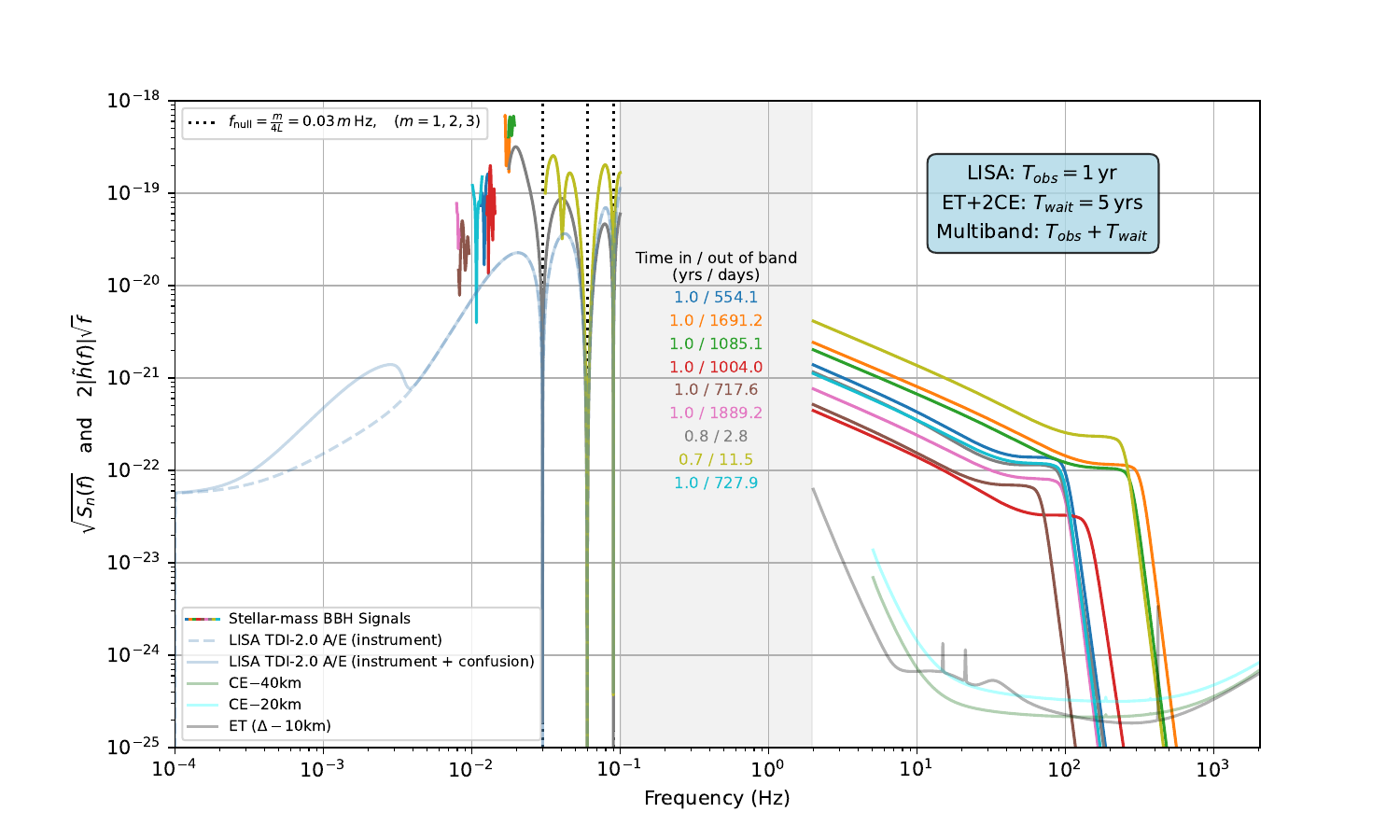}
    \caption{Multiband observations of stellar-mass black hole binaries (sBBHs). This figure illustrates one of the population simulations in Sec.~\ref{sec:population_results}. In this scenario, LISA observes for one year, while the ground-based detectors (ET+2CE) operate for six years ($T_{obs} + T_{wait}$) starting from the beginning of LISA's observation. We focus on coalescence events whose signals enter the ground-based frequency band overlapping with or following LISA's observation, ensuring multiband coverage of systems observed by LISA. The colored lines represent LISA signals ($\rho_{LISA} \ge 3$) modulated by TDI-2 detector response instead of showing characteristic strain. Faint lines depict the square root of PSD (ASD) of TDI-A/E and ET/CE. Most sBBH signals lie above Galactic double white dwarfs (DWD) confusion noise but IMBHBs might remain susceptible. Limited LISA observation time (1 year) causes narrowband clustering near $10^{-2}$ Hz due to small frequency evolution. Amplitude modulation arises from LISA's orbital motion. Two high-frequency signals experience three Michelson TDI null frequencies \citep{Wang:2024alm} during 1-year observation, exiting LISA's band and entering ground-based detectors within 3-12 days before merger. The gray region encompasses all signal timelines. This demonstrates multiband observation feasibility even with LISA's short observing period.}
    \label{fig:multiband_asd}
\end{figure*}

\section{\label{sec:background}Summary of an expected scenario of a multiband observation}

% AHN: Add a section that summarizes the expected population of multi-band source
% A place to give a picture of say how the a signal evolves from LISA -> ground-based detectors
% Add in to the plot TianQin / Taiji / DECIGO
% this is a place to start discussing what you learn from different freuqencies

Fig.~\ref{fig:multiband_asd} illustrates a simulated scenario where LISA observes gravitational wave signals for one year ($T_{obs} = 1$ year). We then consider sources that would enter the sensitive frequency range of ground-based detectors within five years ($T_{wait} = 5$ years) after LISA's observation concludes. This setup demonstrates how space-borne and ground-based detectors could collaborate in a multiband observation strategy. The SNR threshold for LISA is set to 3. It's based on one of the multiband realizations from Sec.~\ref{sec:population_runs}.

The slightly dimmer lines show the amplitude spectral density (ASD) of the detectors that we used in this paper. For LISA, we chose the second generation of time-delay interferometry (TDI-2) \citep{Tinto:2020fcc} power spectral density (PSD) with the confusion noise made by the Galactic double white dwarfs (DWD) \citep{Babak:2021mhe}. In the next-generation ground-based detectors, ET-D was selected as the sensitivity for Einstein Telescope \citep{Hild:2010id}, and the corresponding 20 km and 40 km sensitivities were used for the two Cosmic Explorer \citep{Evans:2023euw}. The multiband sBBH signals above the SNR threshold are highlighted.

To be as realistic as possible, we directly plotted the frequency-domain LISA TDI waveforms in the figure. With only a one-year observation time, most multiband signals will not fully evolve through the LISA frequency band, they are mostly concentrated around $10^{-2}$ Hz, higher than the DWD confusion noise band (around $10^{-3}$ Hz). However, if IMBHBs exist, because of their heavier mass and lower GW frequency, these signals will likely be affected by confusion noise. Most of the sBBH signals span the entire 1-year LISA observation time, but two signals evolved out of the LISA band ($>$ 0.1 Hz) within the first year of LISA observation. These two signals, due to their heavier mass and higher GW radiation efficiency, after leaving the LISA frequency band, entered the 3G detector frequency band in just a few days. The remaining signals spent several years in the frequency gap before entering the ground detector band. The duration of the late-inspiral, merger and ringdown ($\mathcal{O}(s) $) is negligible compared to the other stages.

The area enclosed by the signal and the detector ASD is the SNR in each frequency band. It can be roughly seen that the SNR contributed by the LISA band is negligible compared to the SNR in 3G detectors. However, this does not mean that LISA signals are not important. As mentioned in Sec.~\ref{sec:intro}: (1) The orbital eccentricity and precession spin of LISA sBBHs provide insight into their formation mechanism; (2) Waveform duration of sBBHs in the LISA band is very long, there are a large number of waveform cycles in the data, so the phase information of the waveform can be accurately measured, which is of great help for testing GR and detecting environmental effects. Complementarily, the extremely high SNR in 3G detectors can improve the overall multiband SNR, which is crucial for strong-field tests of gravity.

\section{\label{sec:method}Coherent multiband parameter estimation}

%  Background on parameter estimation for both ground and LISA
%  mathematical formulation
%  -> highlight the differences 
%      -> ground based you can factor sky locatoin into a set of cofficients 
%      -> this is why you can do very fast sky marginalization for ground-based detectors
%  -> explain the connection to how to speed up LISA data anlaysis
%  
% -> highly constrained sky location means that the LISA likelihood can be accurately approximated
% -> by the ML ground-based prediction. (1) in most cases, the posterior is unimodal (2) demonstrate with your mismatch plots, and (3) explain how certain levels of mismatch are appropriate and won't cause bias given the low SNR of the LISA signals
% -> walk through how you are approximate the LISA Likelhood and what you do to map it to the the target one
% -> amplitude / phase rescaling? (maybe mention for future) 

% add an example
% maybe an example with HM for ground-based

This section introduces the new method of this paper in detail: the Bayesian multiband parameter estimation method that keeps the coherence of LISA and 3G detector's waveforms' phase and marginalizes over extrinsic parameters. This new method can overcome the difficulties of existing methods in analyzing LISA sBBHs to some extent, such as difficulties caused by an extremely low SNR, long-lived signal duration, and complex high-dimensional parameter space.

% In the Bayesian parameter estimation of GW signals, likelihood is a key component. If observed data is $d$ and the GW signal can be described by a set of parameters $\bm{\theta}$, then likelihood $p(d | \bm{\theta})$ is the probability of observaing data $d$ given the signal defined by $\bm{\theta}$,

% \begin{equation}
% p(\bm{\theta} | d) = \frac{p(d | \bm{\theta}) \pi(\bm{\theta})}{\int p(d | \bm{\theta}) \pi(\bm{\theta}) d\bm{\theta}}, \label{eq:true_posterior}
% \end{equation}
% where $\pi(\bm{\theta})$ and $p(\bm{\theta} | d)$ are prior and posterior of those parameters, respectively. They represent our knowledge of those parameters before and after the observation.

In the Bayesian parameter estimation of GW signals, the likelihood function $p(d|\bm{\theta})$ plays a central role. Given observed data $d$ and a signal model parameterized by $\bm{\theta}$, the likelihood quantifies the conditional probability of observing the data $d$ under the hypothesis that the signal is described by the parameters $\bm{\theta}$. According to Bayes' theorem, the posterior probability distribution over the parameters is given by:
\begin{equation}
p(\bm{\theta} | d) = \frac{p(d | \bm{\theta}) \pi(\bm{\theta})}{\int p(d | \bm{\theta}) \pi(\bm{\theta}) d\bm{\theta}}, \label{eq:true_posterior}
\end{equation}
where $\pi(\bm{\theta})$ represents the prior distribution encoding our state of knowledge about the parameters before observing the data, and the denominator ensures proper normalization of the posterior distribution. This formulation provides a principled way to update our parameter knowledge from the prior $\pi(\bm{\theta})$ to the posterior $p(\bm{\theta}|d)$ in light of the observed data.

The matched-filtering technique \citep{Allen:2005fk} underpins template-based GW detection. Rooted in the Neyman-Pearson lemma \citep{Neyman:1933wgr,Finn:1992wt,Owen:1998dk,jaranowski2009analysis}, which identifies the optimal hypothesis test via the likelihood ratio between signal and noise hypotheses. Expressed for GW data, this ratio becomes

\begin{subequations}
\label{eq:group}
\begin{align}
\ln \mathcal{L}(\bm{\theta}) &= \ln \frac{p(d | \text{signal}, \bm{\theta})}{p(d | \text{noise})} \\
&= \langle d | h(\bm{\theta}) \rangle - \frac{1}{2} \langle h(\bm{\theta}) | h(\bm{\theta}) \rangle,
\end{align} \label{eq:loglr}
\end{subequations}
with

\begin{equation}
\left\langle a | b\right\rangle=4 \Re \int_{f_{\text {min}}}^{f_{\text {max}}} \frac{\tilde{a}(f) \tilde{b}^{*}(f)}{S_{n}(f)} \mathrm{~d} f,
\label{eq:inner_product}
\end{equation}
here $p(d | \text{signal}, \bm{\theta})$ is same as $p(d | \bm{\theta})$. This ratio represents the relative possibility under two hypotheses (at the parameter $\bm{\theta}$). $S_{n}$ is the one-sided power spectral density of the detector noise, and ``*" denotes the complex conjugate, ``$\sim$" denotes the Fourier transform. Since the signal itself has intrinsic and extrinsic parameters, in order to eliminate the uncertainty caused by the parameters, what is ideally used for signal detection is the likelihood ratio marginalized over all parameters, which is the Bayes factor between the hypotheses ``contain signal" and ``no signal". In Bayesian parameter estimation of GW signals, the log-likelihood ratio Eq.~(\ref{eq:loglr}) is usually used to sample the posterior distribution. However, it is worth noting that since the noise likelihood $p(d | \text{noise})$ is a constant given the data, the likelihood ratio is essentially proportional to the signal's likelihood function.

Ideally, in multiband parameter estimation, we need to calculate the joint multiband likelihood of LISA and ground-based detectors during the sampling,

\begin{equation}
\begin{aligned}
\ln \mathcal{L}_{\text{joint}}(\bm{\theta}) = & \sum_{c \in \mathcal{C}_{\text{LISA}}} \left( \langle d_c | h_c(\bm{\theta}_\text{LISA}) \rangle - \frac{1}{2} \langle h_c(\bm{\theta}_\text{LISA}) | h_c(\bm{\theta}_\text{LISA}) \rangle \right) \\
& + \sum_{d \in \mathcal{D}_{\text{3G}}} \left( \langle d_d | h_d(\bm{\theta}_\text{3G}) \rangle - \frac{1}{2} \langle h_d(\bm{\theta}_\text{3G}) | h_d(\bm{\theta}_\text{3G}) \rangle \right),
\end{aligned}
\label{eq:multiband_loglr}
\end{equation}
$c \in \mathcal{C}_{\text{LISA}}$ means loop over all LISA TDI channels, and $d \in \mathcal{D}_{\text{3G}}$ means loop over all ground-based detectors. $\bm{\theta}_\text{LISA}$ and $\bm{\theta}_\text{3G}$ are equivalent, but in different coordinate systems, due to different conventions in LISA and ground-based analysis. After we have chosen a specific frame, we can convert them into the same $\bm{\theta}$. But this is technically challenging, and no previous multiband paper completely realized and calculated it in the Bayesian analysis. Upon completion of this work, we became aware of related work by \citep{Linley:2022ndb}, which also proposed a joint likelihood, but with a much simplified treatment (such as a sky-averaged LISA response, no coordinate transform and noise-free). In the following subsections, we will introduce how we can calculate it accurately and efficiently.

\subsection{\label{subsec:coordinate}Coordinate Systems and Transforms in Multiband Analysis}

\begin{figure*}
    \centering
    \vspace*{-0.5cm}
    \includegraphics[width=1.0\textwidth]{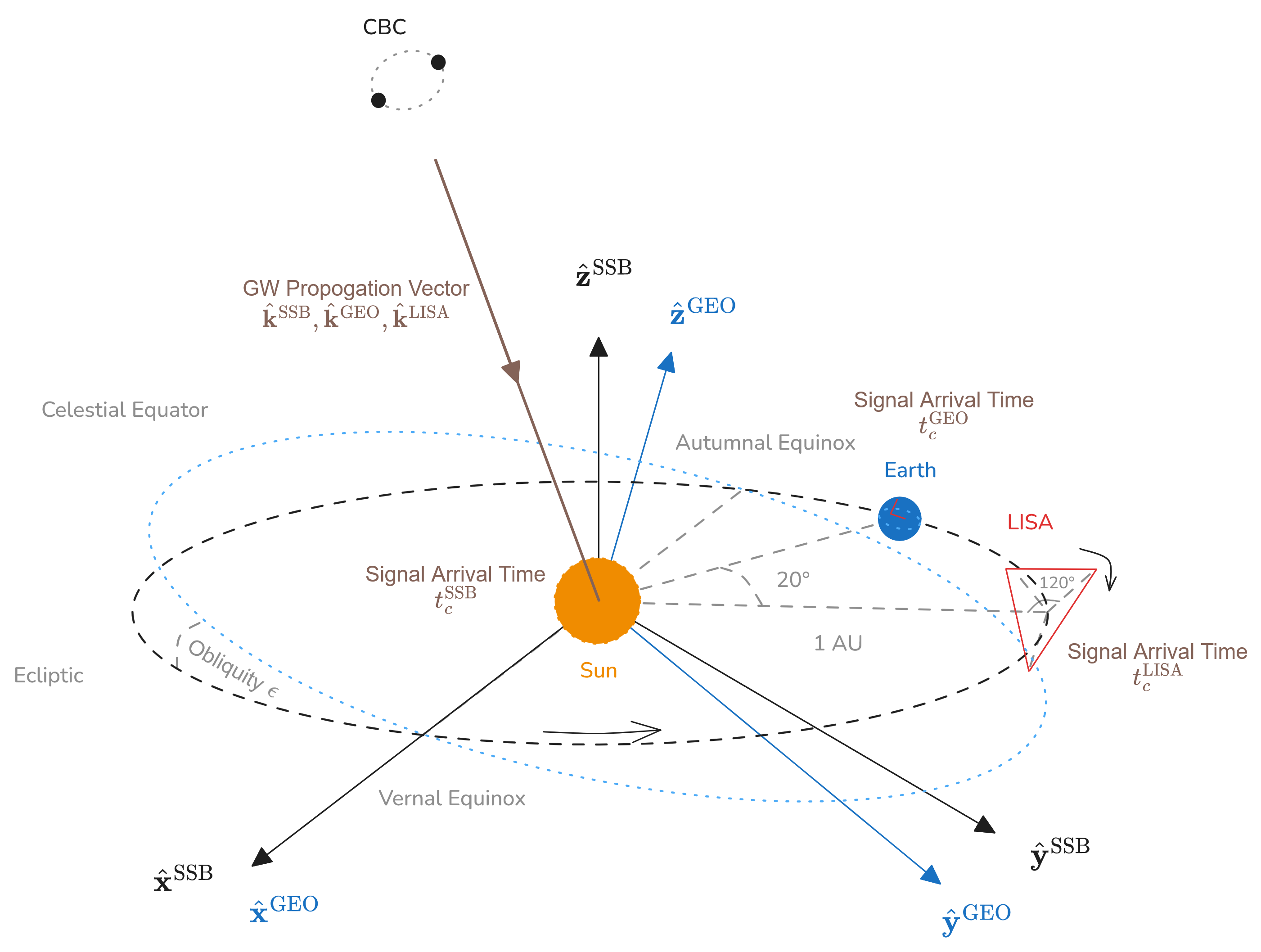}
    \caption{Schematic depiction of orbits of the Earth and LISA, also with coordinate systems used by LISA and ground-based detectors, not to scale. Note that we don't show polarization angle in each frames for simplicity.}
    \label{fig:frame}
\end{figure*}

Strictly speaking, in the multiband analysis of LISA+3G detectors, we need to consider the transformation between five coordinate systems shown in Fig.~\ref{fig:frame}: (1) the source frame of the GW source; (2) the Solar System Barycenter (SSB) frame describing the motion of LISA and the Earth in the solar system; (3) the LISA frame bound to the three satellites of LISA; (4) the geocentric (GEO) frame commonly used for ground-based detectors; (5) when calculating GW strain using ground-based detectors, it is essential to account for the detector frame's position and orientation.

Some extrinsic parameters are frame dependent: (1) the signal arrival time at the origin of the coordinate system (we choose the time $t_{c}$ when the merge signal arrives); (2) the unit propagation vector $\mathbf{\hat{k}}$ of the signal (corresponding to the location of the GW source in the frame); (3) the polarization angle $\psi$ of the signal relative to the frame. To put all signals among different detectors in the same extrinsic parameter space in multiband Bayesian parameter estimation, we need to perform frame transformations in real-time during the sampling. We can achieve fast transformations between different frames based on rotation matrices, which we detail in the appendix~\ref{sec:appendix_ssb_lisa_geo}.

In this paper, we perform multiband Bayesian parameter estimation in the geocentric frame for both the LISA and ET+2CE detectors; the reason is that the SNR of the signal is dominated by 3G detectors, and we can use information from ground-based detectors to marginalize parameter space, which will be discussed later.

\subsection{\label{subsec:det_strain}Detector Strain of Ground-based GW Detectors and LISA}

The GW strain recorded by a certain ground-based GW detector $d$ can be expressed by the following formula (for simplicity, here we only consider the dominant mode of the GW, namely the (2, $\pm$2) mode). We can separate its dependence on intrinsic parameters and extrinsic parameters,

\begin{equation}
\begin{split}
&\tilde{h}_d\left(f; \bm{\theta}_{\text{int}}, \mathbf{\hat{k}}^{\text{GEO}}, \psi^{\text{GEO}}, t_c^{\text{GEO}}, \phi_{\text{ref}}, D_L\right) \\
=& \frac{1}{D_L} \sum_{p \in \{+, \times\}} \tilde{h}_{22}^p\left(f; \bm{\theta}_{\text{int}}\right) F_d^p\left(\mathbf{\hat{k}}^{\text{GEO}}, \psi^{\text{GEO}}\right) \\
&\times e^{-i 2 \pi f t_d\left(t_c^{\text{GEO}}, \mathbf{\hat{k}}^{\text{GEO}}\right)} e^{i 2 \phi_{\text{ref}}} \label{eq:xg_det_strain}
\end{split}
\end{equation}
with

\begin{equation}
\tilde{h}_{22}^p\left(f; \bm{\theta}_{\text{int}}\right) := \tilde{h}_{22}^p\left(f; \bm{\theta}_{\text{int}}, D_L = 1, \phi_{\text{ref}} = 0\right),
\label{eq:xg_intrinsic_strain}
\end{equation}
where $\mathbf{\hat{k}}^{\text{GEO}}$ is the unit propagation vector of the GW signal in the geocentric frame (equivalent to the right ascension and declination of the corresponding source location of the GW). The polarization angle $\psi^{\text{GEO}}$ reflects the rotation angle of the GW source frame with respect to the geocentric frame along the unit propagation vector $\mathbf{\hat{k}}^{\text{GEO}}$. $t_c^{\text{GEO}}$ is the time when the GW signal arrives at the origin of the geocentric frame (which we can choose as the time corresponding to the merger of the CBC signal). $\phi_{\text{ref}}$ is the orbital phase of the binary system at a reference frequency $f_{\text{ref}}$. $D_{L}$ is the luminosity distance between the GW source and the origin of the geocentric frame. $\tilde{h}_{22}^p$ is the GW that only depends on the intrinsic parameters (distance is normalized, reference phase is zero), $F_d^p$ is the antenna response function of the GW detector $d$ to the GW polarization mode $p$ (only the long-wavelength approximation is considered here, without considering the effect of the Earth's rotation, so it does not contain time). The $e^{-i 2 \pi f t_d}$ reflects the phase shift of the GW signal caused by the propagation delay of the signal from the origin of the geocentric frame to the detector. The $e^{i 2 \phi_{\text{ref}}}$ reflects the effect of the binary orbital phase $\phi_{\text{ref}}$ on the phase of the GW signal.

The inner product of the GW detector data $d$ and the GW signal template $\tilde{h}_d$ is

\begin{equation}
\begin{split}
\langle d | h \rangle &= \frac{1}{D_L} \, \Re \biggl[e^{-i 2 \phi_{\text{ref}}} \sum_{p,d} F_d^p\left(\mathbf{\hat{k}}^{\text{GEO}}, \psi^{\text{GEO}}\right) \\
&\quad \times z_{22}^d\left(t_d\left(t_c^{\text{GEO}}, \mathbf{\hat{k}}^{\text{GEO}}\right); \bm{\theta}_{\text{int}}\right) \biggr]
\label{eq:inner_product_dh}
\end{split}
\end{equation}
with

\begin{equation}
z_{22}^d\left(t; \bm{\theta}_{\text{int}}\right) := 4 \int_{f_{\text{min}}}^{f_{\text{max}}} df \, \frac{\tilde{d}_d(f) \left[ \tilde{h}_{22}^p\left(f; \bm{\theta}_{\text{int}}\right) \right]^* e^{i 2 \pi f t}}{S_n(f)}.
\label{eq:complex_snr}
\end{equation}

Eq.~(\ref{eq:complex_snr}) is the complex matched filtering SNR time series of the GW signal polarization $p$ with data $d$, and Eq.~(\ref{eq:inner_product_dh}) contains its SNR peak corresponding to the arrival time of the merger signal. The inner product of the template itself is

\begin{equation}
\begin{split}
&\langle h | h \rangle \\
=&\frac{1}{D_L^2} \sum_{p,p',d} C_{22}^{pp'd}\left(\bm{\theta}_{\text{int}}\right) F_d^p\left(\mathbf{\hat{k}}^{\text{GEO}}, \psi^{\text{GEO}}\right) F_d^{p'}\left(\mathbf{\hat{k}}^{\text{GEO}}, \psi^{\text{GEO}}\right)
\label{eq:inner_product_hh}
\end{split}
\end{equation}
with

\begin{equation}
C_{22}^{pp'd}(\bm{\theta}_{\text{int}}) = 4 \int_{f_{\text{min}}}^{f_{\text{max}}} df \, \frac{\tilde{h}_{22}^p(f; \bm{\theta}_{\text{int}}) \left[ \tilde{h}_{22}^{p'}(f; \bm{\theta}_{\text{int}}) \right]^*}{S_n(f)},
\label{eq:cov_matrix_inner_product}
\end{equation}
where $C_{22}^{pp'd}$ is the inner product covariance matrix between the two dominant polarization modes of the GW signal.

Compared to the response of ground-based detectors to GW signals, the situation for LISA is much more complex. First, the duration of the signals observed in its mHz band is measured in months or years, which is comparable to LISA's orbital period around the Sun. Moreover, LISA itself has its own rotation around its axis. LISA's revolution and rotation lead to time-varying detector response functions and Doppler effects. Furthermore, since the wavelength of signals in the LISA band is comparable to the arm length of LISA itself, the long-wavelength approximation commonly used in ground-based detector analysis becomes invalid. In addition, LISA needs to use TDI to eliminate laser frequency noise \citep{Tinto:2020fcc}. In this paper, the LISA waveform and response calculation uses the $\texttt{BBHx}$ \citep{Katz:2020hku} waveform (although it supports GPU calculation, we only use CPU in this work) through its $\texttt{PyCBC}$ plugin \footnote{The \texttt{PyCBC} waveform plugin of \texttt{BBHx} \url{https://github.com/gwastro/BBHX-waveform-model}},

\begin{equation}
\tilde{h}^{\text{A,E,T}}_{22}(f, t_{22}^{\text{LISA}}(f)) = \mathcal{T}^{\text{A,E,T}}(f, t_{22}^{\text{LISA}}(f)) \tilde{h}_{22}(f)
\label{eq:lisa_aet_strain}
\end{equation}
with

\begin{equation}
t_{22}^{\text{LISA}}(f) = t_{\text{c}}^{\text{LISA}} - \frac{1}{2\pi} \frac{d\phi_{22}(f)}{df},
\label{eq:lisa_tf_track}
\end{equation}
here $\tilde{h}^{\text{A,E,T}}_{22}$ represents the GW strain in the three quasi-orthogonal TDI channels (A, E, T) of LISA \citep{Tinto:2020fcc}. $t_{22}^{\text{LISA}}(f)$ is the time-frequency track of the dominant mode of GW. $\mathcal{T}^{\text{A,E,T}}$ is the response transfer function of the three TDI channels, similar to ground-based detectors, it is also a function of extrinsic parameters, but they're omitted here for simplicity. $t_{\text{c}}^{\text{LISA}}$ is the time when the GW merger signal arrives at the centroid of LISA in the LISA frame. $\phi_{22}(f)$ is the function of the dominant mode phase of GW with respect to frequency $f$.

Similarly to the calculation of the likelihood ratio for ground-based gravitational wave detectors discussed earlier, we use Eq.~(\ref{eq:lisa_aet_strain}) (instead of Eq.~(\ref{eq:xg_det_strain})) as the waveform to calculate those inner products. In addition, we need to use the PSD of LISA's A, E, and T channels as the corresponding $S_{n}(f)$. In the specific implementation of \texttt{PyCBC}, the contribution of Galactic white dwarf binaries' confusion noise to the TDI PSD is also added (see Figure.~\ref{fig:multiband_asd}, based on the phenomenological fitting from \citep{Babak:2021mhe}).

\subsection{\label{subsec:margin_lr_imp_sample}Marginalized Likelihood and Importance Sampling}

To simplify the complex parameter space, we can use Bayesian parameter marginalization \citep{Thrane:2018qnx} to reduce the dimensionality of the parameter space. In parameter estimation, extrinsic parameters can be marginalized:

\begin{equation}
\bar{L}(d | \bm{\theta}_{\text{int}}) = \int p\left(d | \bm{\theta}_{\text{int}}, \bm{\theta}_{\text{ext}}\right) \pi(\bm{\theta}_{\text{ext}}) \, d\bm{\theta}_{\mathrm{ext}}, \label{eq:margin_likelihood}
\end{equation}
where $p\left(d | \bm{\theta}_{\text{int}}, \bm{\theta}_{\text{ext}}\right)$ is the probability of obtaining the observed data given the intrinsic parameters $\bm{\theta}_{\text{int}}$ and extrinsic parameters $\bm{\theta}_{\text{ext}}$, which is the likelihood function ${L}$. $\pi(\bm{\theta}_{\text{ext}})$ is the prior distribution of the extrinsic parameters $\bm{\theta}_{\text{ext}}$, which reflects the expectations of people about this quantity before analyzing the data. When the likelihood function ${L}$ is weighted by the prior of the extrinsic parameters and integrated over them, the marginalized likelihood function of Eq.~({\ref{eq:margin_likelihood}}) is obtained, which is only a function of the intrinsic parameters (such as mass and spin), and its dimension of parameter space is significantly reduced. If we obtain this marginalized likelihood function, and use it to sample the intrinsic parameter space, the parameter estimation problem will be greatly simplified.

According to Bayes' theorem, the conditional posterior of the extrinsic parameters (the posterior probability given observed data and intrinsic parameters) is

\begin{equation}
p(\bm{\theta}_{\text{ext}} | d, \bm{\theta}_{\text{int}}) = \frac{p(d | \bm{\theta}_{\text{int}}, \bm{\theta}_{\text{ext}}) \pi(\bm{\theta}_{\text{ext}})}{\bar{L}(d | \bm{\theta}_{\text{int}})}, \label{eq:conditional_posterior}
\end{equation}
where the denominator on the right side of the formula is the likelihood function after marginalizing the extrinsic parameters, that's Eq.~(\ref{eq:margin_likelihood}). Once we get this marginalized likelihood function, the conditional posterior of the extrinsic parameters can also be easily obtained.

However, in GW data analysis, the marginalized likelihood function of Eq.~({\ref{eq:margin_likelihood}}) is generally difficult to calculate analytically, because the likelihood function $L$ itself is obtained through numerical calculation. In this case, we need to use importance sampling \citep{Roulet:2024hwz,Nitz:2024nhj}, which is a Monte Carlo numerical integration method used to estimate the expectation or integral. The key to importance sampling is that when it is difficult or inefficient to sample directly from the target distribution $p(x)$, a proposal distribution $q(x)$ that is easier to sample can be selected for sampling. The weight of each sample $x$ is calculated as the ratio of the target distribution to the proposal distribution $\frac{p(x)}{q(x)}$, and the expectation or integral of the original distribution is approximated (the number of samples is finite) by weighted averaging,

\begin{equation}
\begin{aligned}
\mathbb{E}_p[f(x)] &= \int f(x)p(x)\,dx = \int f(x)\frac{p(x)}{q(x)}q(x)\,dx \\
&= \mathbb{E}_q\left[\frac{p(x)}{q(x)}f(x)\right] \simeq \frac{1}{N} \sum_{i=1}^N w(x_i) f(x_i),
\end{aligned} \label{eq:importance_sampling}
\end{equation}
where $w(x_i) = \frac{p(x_i)}{q(x_i)}$ is the importance weight for sample $x_i$, this $x_i$ is drawn from the proposal distribution $q(x)$. As we can see here, it transforms the expectation of $f(x)$ over $p$ into the expectation of $\frac{p(x)}{q(x)}f(x)$ over $q$ instead. When that proposal distribution $q(x)$ is similar to the target distribution $p(x)$, this importance sampling technique can achieve much lower variance of the integration for a given number of samples $N$, compared to totally random Monte Carlo numerical integration.

We can apply importance sampling Eq.~(\ref{eq:importance_sampling}) to the original form of marginalized likelihood Eq.~(\ref{eq:margin_likelihood}) as following

\begin{subequations}
\label{eq:group_margin_logl}
\begin{align}
\bar{L}(d | \bm{\theta}_{\text{int}})
&= \int p\left(d | \bm{\theta}_{\text{int}}, \bm{\theta}_{\text{ext}}\right) \pi(\bm{\theta}_{\text{ext}}) \, d\bm{\theta}_{\mathrm{ext}} \label{eq:direct_a} \\
&= \int \frac{p\left(d | \bm{\theta}_{\text{int}}, \bm{\theta}_{\text{ext}}\right) \pi(\bm{\theta}_{\text{ext}})}{q(\bm{\theta}_{\text{ext}})} q(\bm{\theta}_{\text{ext}}) \, d\bm{\theta}_{\mathrm{ext}} \label{eq:direct_b} \\
&= \mathbb{E}_{q(\bm{\theta}_{\text{ext}})} \left[ \frac{p\left(d | \bm{\theta}_{\text{int}}, \bm{\theta}_{\text{ext}}\right) \pi(\bm{\theta}_{\text{ext}})}{q(\bm{\theta}_{\text{ext}})} \right] \label{eq:direct_c} \\
&\simeq \frac{1}{N} \sum_{i=1}^N \frac{p\left(d | \bm{\theta}_{\text{int}}, \bm{\theta}_{\text{ext}}^i\right) \pi(\bm{\theta}_{\text{ext}}^i)}{q(\bm{\theta}_{\text{ext}}^i)} \label{eq:direct_d} \\
&= \frac{1}{N} \sum_{i=1}^N w_i, \quad  \label{eq:direct_e}
\end{align}
\end{subequations}
we can see that the extrinsic marginalized likelihood function $\bar{L}(d | \theta_{\text{int}})$ can be approximated by the average of the importance sampling weights $w_i$. These $w_i$ are the ratios of the conditional posterior distribution (unnormalized) to the proposal distribution $q(\bm{\theta}_{\text{ext}})$. The value of the conditional posterior (unnormalized) is easy to obtain, and the proposal distribution is also defined by ourselves, which is also easy to calculate. Therefore, the calculation of the entire marginalized likelihood function becomes feasible. The next question is how to choose a suitable proposal distribution $q(\bm{\theta}_{\text{ext}})$ in the GW data analysis. If not chosen properly, it will lead to a large variance in the final estimated marginalized likelihood and reduce computational efficiency. Theoretically, the efficiency can be optimal when the proposal distribution is proportional to the target distribution, in this case, the unnormalized conditional posterior distribution $p(d | \bm{\theta}_{\text{int}}, \bm{\theta}_{\text{ext}}) \pi(\bm{\theta}_{\text{ext}})$. We will describe the proposal explicitly in the next section.

Marginalized likelihood can be used to sample the intrinsic parameter space more efficiently. After sampling, we can obtain the posterior $p(\bm{\theta}_{\text{int}} | d)$ of the intrinsic parameters. For each intrinsic sample $\bm{\theta}^{k}_{\text{int}}$, it has a set of importance weights $\{\bm{\theta}^i_{\text{ext}}, w_i^{k}\}_{i=1}^N$. The normalized importance weight will come close to the posterior value at the extrinsic parameter $\bm{\theta}^i_{\text{ext}}$, when the number of marginalization samples $N$ is large (law of large numbers),

\begin{equation}
\begin{aligned}
\bar{w}_i = \frac{w_i}{\sum_i w_i} \approx p(\bm{\theta}^i_{\text{ext}} | d, \bm{\theta}_{\text{int}}),
\end{aligned} \label{eq:importance_weight_normalized}
\end{equation}
which means we can use the normalized weight $\bar{w}_i$ to draw posterior samples from prior for extrinsic parameters. In a small neighborhood \( d\bm{\theta}_{\text{ext}} \) around \(\bm{\theta}^i_{\text{ext}}\), as \( N \to \infty \), we have the following: (1) the proportion of samples (drawn from the proposal distribution) falling into this region is \(\propto q(\bm{\theta}_{\text{ext}}) d\bm{\theta}_{\text{ext}}\); (2) the importance weight of each sample is \(\propto p(\bm{\theta^i}_{\text{ext}} | d, \bm{\theta}_{\text{int}})/q(\bm{\theta}_{\text{ext}})\); (3) the total weight in this region is \(\propto p(\bm{\theta}^i_{\text{ext}} | d, \bm{\theta}_{\text{int}})/q(\bm{\theta}_{\text{ext}}) \cdot q(\bm{\theta}_{\text{ext}}) d\bm{\theta}_{\text{ext}} = p(\bm{\theta}^i_{\text{ext}} | d, \bm{\theta}_{\text{int}}) d\bm{\theta}_{\text{ext}}\). In this way, we can get the full posterior.

\subsection{\label{subsec:multiband_margin}Multiband Marginalization of Extrinsic Parameters}

\begin{figure}[h]
	\includegraphics[width=0.49\textwidth]{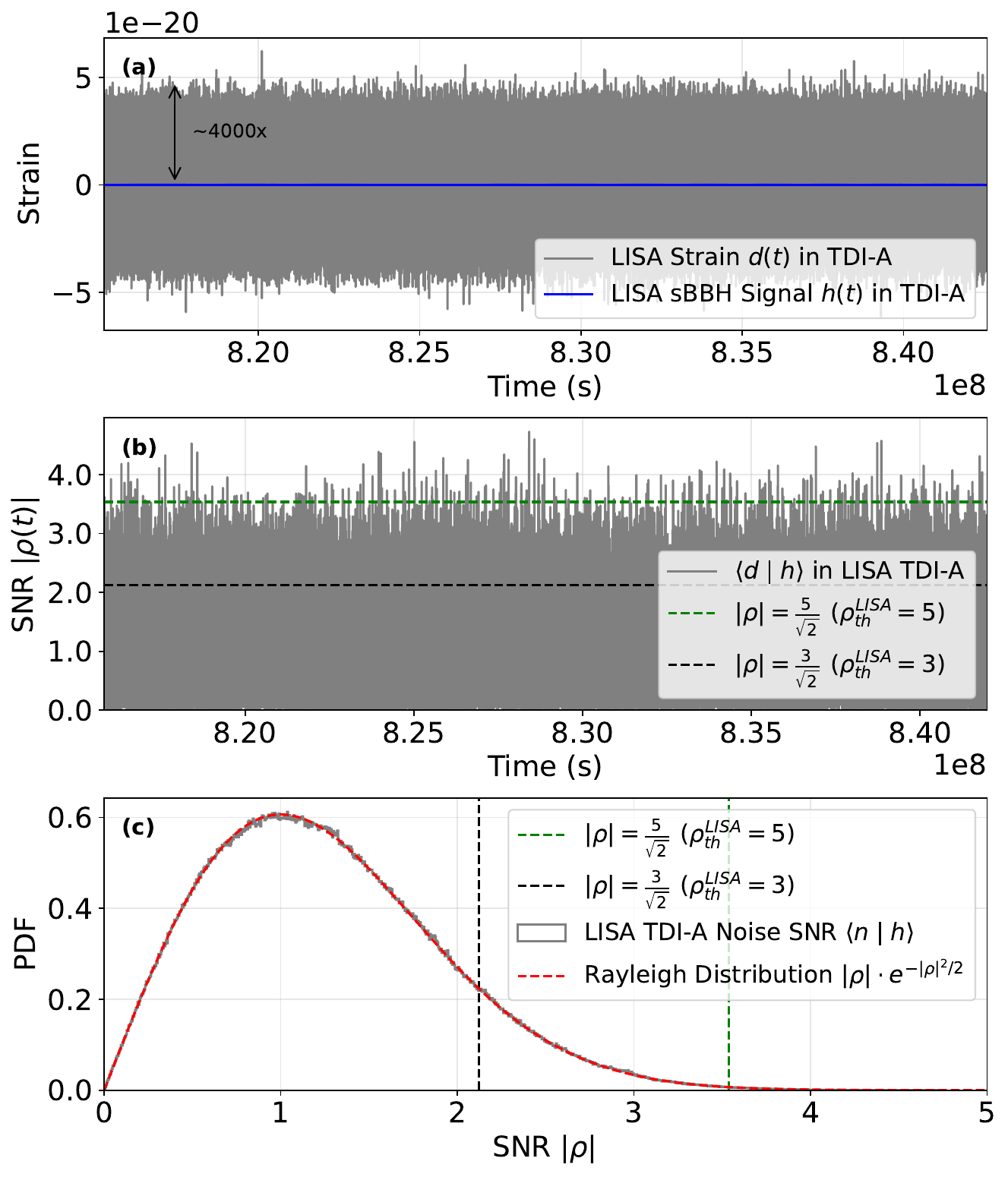}
    \caption{LISA's sBBH signal in the TDI channel. Panel (a) shows a typical sBBH signal and the total strain, including detector noise in the TDI-A channel. The signal is very weak, about 4000 times weaker than LISA noise in time-domain amplitude. Panel (b) shows the modulus of complex matched-filtering SNR time series of the data $d(t)$ using the signal $h(t)$ as a template, which also reflects likelihood ratio $\mathcal{L}$ of $t_c$, considering $\rho = \langle d | h \rangle / \sqrt{\langle h | h \rangle} \approx \sqrt{2 \ln \mathcal{L}}$. Note that we omit the normalization term $\sqrt{\langle h | h \rangle}$ in the legend of figure for simplicity. When we set the LISA SNR threshold to 5 or 3, the SNR corresponding to a single TDI channel is roughly the overall threshold divided by $\sqrt{2}$. This is because the TDI-A and TDI-E channels almost equally share the GW signal. The sBBH signal is already below the SNR of LISA's noise itself. The limit of the LISA SNR in the previous multiband analysis was 5, and the reason for this is easily seen in the figure. Panel (c) is the PDF distribution of the same SNR time series, its amplitude still follows the Rayleigh distribution, which is consistent with the situation of ground-based detectors \citep{Wu:2022pyg}.}
    \label{fig:strain_snr_sbbh_lisa}
\end{figure}

It is difficult to directly analyze sBBH signals in LISA itself (low SNR and so many local maxima), as shown in Figure~\ref{fig:strain_snr_sbbh_lisa}, projecting the data $d(t)$ onto the signal manifold $h(t)$, there are so many local maxima. Even if we use matched filtering, the noise's SNR is still likely higher than the signal's SNR. That means we need to use the information provided by ground-based GW detectors to extract sBBH signals in LISA.

For different extrinsic parameters, we have different marginalization approaches: (1) for luminosity distance, we can use the universal scaling relation to build a look-up table of inner product with different luminosity distance, and numerically integrate it; (2) for orbital phase, we can marginalize analytically; (3) for arrival time and sky position, we can use the importance sampling described in Sec.~\ref{subsec:margin_lr_imp_sample}.

\begin{figure}[h]
	\includegraphics[width=0.49\textwidth]{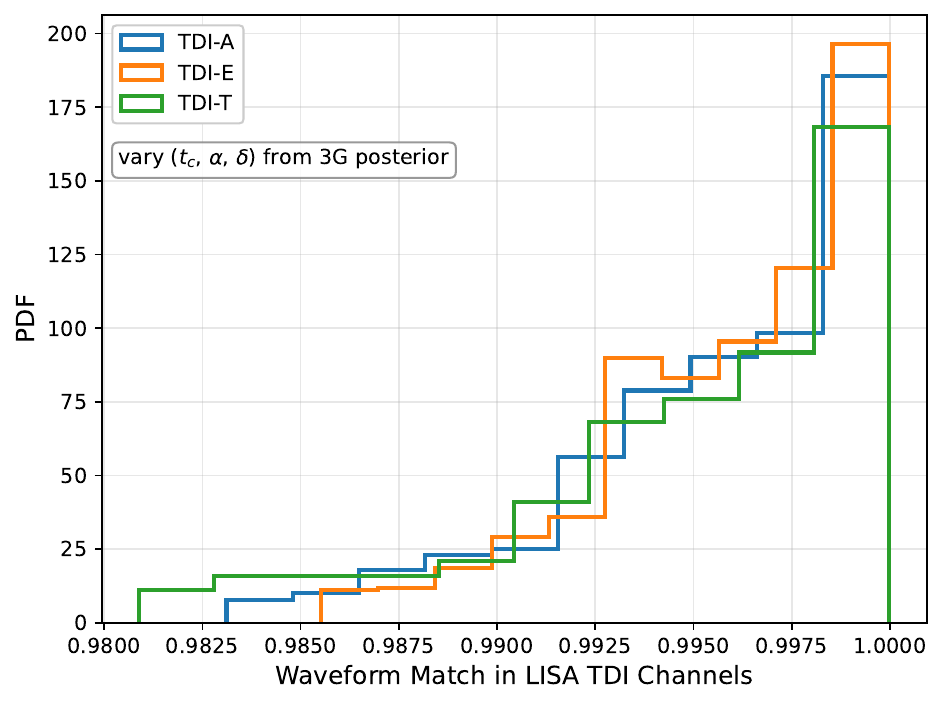}
    \caption{The effect of extrinsic parameters ($t_c, \alpha, \delta$) on LISA waveforms. By varying the posterior of ($t_c, \alpha, \delta$) from ET+2CE, we calculated the match of the corresponding LISA waveforms relative to the true waveform. The waveform mismatch is within 2\%. Considering the extremely low SNR in LISA, the mismatch is negligible.}
    \label{fig:match_tdi_waveform}
\end{figure}

In Fig.~\ref{fig:match_tdi_waveform}, the arrival time and localization parameters of the corresponding LISA waveform are adjusted (using 3G posterior values), and the distribution of the match calculation (inner product after time and phase maximization) is carried out with the actual LISA injection signal. The waveforms of the three TDI channels exhibit minimal variation over time and location, and the mismatch and the loss of SNR are within 2\%. In importance sampling, we can not only use the posterior distribution of ($t_c, \alpha, \delta$) of ET+2CE as the proposal distribution for the entire multiband analysis, but we can also use the same ($t_c, \alpha, \delta$) corresponding to the maximum SNR in 3G detectors for all importance samples, which can greatly improve the efficiency of the calculation of the multiband likelihood.

\subsubsection{\label{subsubsec:margin_dl}Marginalization over Luminosity Distance}
% \paragraph{Marginalization over Luminosity Distance}

For both the LISA and the 3G detectors, the dependence of any GW signal on $D_{L}$ is universal; that is, the signal amplitude in the detector frame is inversely proportional to $D_{L}$. As can be seen from Eq.~({\ref{eq:loglr}), the likelihood ratio depends on the calculation of 2 inner products,

\begin{subequations}
\label{eq:group_margin_dl}
\begin{align}
\overline{\mathcal{L}}_{D_L} &= \int \pi(D_L) \exp\left( \frac{\langle d | h_{\text{ref}} \rangle}{D_L/D_L^{\text{ref}}} - \frac{\langle h_{\text{ref}} | h_{\text{ref}} \rangle}{(D_L/D_L^{\text{ref}})^2} \right) \, dD_L \\
&\approx \sum_{i=1}^{N_{D_L}} w_i \cdot \exp\left( \frac{\langle d | h_{\text{ref}} \rangle}{D_{L}^{i}/D_L^{\text{ref}}} - \frac{\langle h_{\text{ref}} | h_{\text{ref}} \rangle}{(D_L^{i}/D_L^{\text{ref}})^2} \right), \label{eq:loglr_margin_dl}
\end{align}
\end{subequations}
where the weights of $D_{L}$ are
$w_i = \int_{D_{L}^{i} - \Delta_{D_L}/2}^{D_{L}^{i} + \Delta_{D_L}/2} \pi(D_L) \, dD_L$. Here, $D_{L}^{\text{ref}}$ is the reference luminosity distance; $h_\text{ref}$ is the signal strain at the reference distance. We divide the distance prior into $N_{D_L}$ segments, each width is $\Delta_{D_L}$. $D_{L}^{i}$ is the mean value of $D_{L}$ for $i$-th segment, the weight $w_i$ is the integral of the prior in each segment. Therefore, we can calculate the values of these two inner products in the likelihood ratio within a given $D_{L}$ prior distribution range as a two-dimensional look-up table for the following numerical integration \citep{Thrane:2018qnx}.

\subsubsection{\label{subsubsec:margin_phi}Marginalization over Orbital Phase}
% \paragraph{Marginalization over Orbital Phase}

If we only consider the dominant mode, the orbital phase $\phi_{\text{ref}}$ (at the reference frequency $f$, we simplify $\phi_{\text{ref}}$ as $\phi$ below) can be analytically marginalized \citep{Veitch:T1300326v1}, regardless of the types of detector,

\begin{subequations}
\label{eq:group_margin_phi}
\begin{align}
&\ln \overline{\mathcal{L}}_{\phi} \\
=& \ln \left[ \int_0^{2\pi} \mathcal{L}(\phi) \frac{d\phi}{2\pi} \right] \\
=& \ln \left[ \int_0^{2\pi} \exp\left( \text{Re}\langle d|h(\phi) \rangle - \frac{1}{2}\langle h|h \rangle \right) \frac{d\phi}{2\pi} \right] \\
=& -\frac{1}{2}\langle h|h \rangle + \ln \left[ \int_0^{2\pi} \exp\left( \text{Re}\left( \langle d|h_0 \rangle e^{i\phi} \right) \right) \frac{d\phi}{2\pi} \right] \\
=& -\frac{1}{2}\langle h|h \rangle + \ln \left[ \int_0^{2\pi} \exp\left( |\langle d|h_0 \rangle| \cos(\phi + \arg \langle d|h_0 \rangle) \right) \frac{d\phi}{2\pi} \right] \\
=& -\frac{1}{2}\langle h|h \rangle + \ln \left[ \int_0^{2\pi} \exp\left( |\langle d|h_0 \rangle| \cos \phi' \right) \frac{d\phi'}{2\pi} \right] \\
=& -\frac{1}{2}\langle h|h \rangle + \ln I_0(|\langle d|h \rangle|),
\label{eq:loglr_margin_phi}
\end{align}
\end{subequations}
where $I_0$ is the modified Bessel function of the first kind. For the higher mode cases, the dependence of each mode is different with the orbital phase, we cannot pull out a common overall phase factor like this. Higher modes multiband marginalization over orbital phase will be investigated in the near future.

\subsubsection{\label{subsubsec:margin_tc_sky}Marginalization over Arrival Time and Localization}
% \paragraph{Marginalization over Arrival Time and Localization}

For extrinsic parameters ($t_{c}$, $\alpha$, $\delta$), we can marginalize using the SNR time series from ground-based GW detectors \citep{Roulet:2024hwz}. Ground-based GW detector networks mainly use triangulation based on the time difference between the signal arrival time at each detector \citep{Fairhurst:2017mvj}. The peak of the SNR time series is a good estimate of the arrival time of the GW signal at each detector. It is best to consider the correlation between them when marginalizing these three extrinsic parameters. From the waveform match in Fig.~\ref{fig:match_tdi_waveform}, the result of the multiband localization should be similar to that of the 3G only case. Therefore, we can utilize the SNR time series from the 3G detectors as a proposal distribution to perform importance-sampling-based marginalization of the multiband likelihood ratio over the parameters ($t_{c}$, $\alpha$, $\delta$). When we select a ground-based detector as the reference detector $d_{0}$, the signal time delay of the remaining detectors $d_{j}$ relative to it is

\begin{subequations}
\label{eq:group_tau}
\begin{align}
\delta t_{d_j} &= \frac{(\hat{r}_{d_j} - \hat{r}_{d_0}) \cdot \hat{\mathbf{k}}^{\text{GEO}}}{c}, \\
\tau_d &= \text{round}\left( \frac{\delta t_d}{\Delta_{\rho}} \right), \label{eq:tc_time_delay}
\end{align}
\end{subequations}
where $r_{d}$ is the detector position vector in the geocentric frame, $\Delta_{\rho}$ is the time resolution of the detector's SNR time series and $\tau_{d}$ is the discretized time delay consistent with the precision of the SNR time series. By iterating through all ground-based detectors, we can obtain a series of discretized time delays $\tau = \{\tau_{d_0}, \tau_{d_1}, \mathellipsis\}$. this set of $\tau$ corresponds to the unit propagation vectors $\hat{\mathbf{k}}^{\text{GEO}}$ of the GW signal. We can establish a mapping between GW localization and time delay,

\begin{equation}
D(\tau) = 
\left\{
(t_c, \hat{\mathbf{k}}^{\text{GEO}})
\,\middle|\,
\begin{aligned}
& |t_{d_0}(t_c, \hat{\mathbf{k}}^{\text{GEO}}) - \tau_{d_0}| < \frac{\Delta_{\rho}}{2}, \\
& |\delta t_d(\hat{\mathbf{k}}^{\text{GEO}}) - \tau_d| < \frac{\Delta_{\rho}}{2} \quad \forall d \neq d_0
\end{aligned}
\right\}, \label{eq:tau_sky_mapping}
\end{equation}
this mapping needs to satisfy two precision conditions: (1) for the reference detector, the absolute error between the exact value of the time delay relative to the Earth's center $t_{d_0}$ and its discrete value $\tau_{d_0}$ should be within an SNR time series precision $\Delta_{\rho} / 2$; (2) the absolute error between the exact value of the time delay $\delta_{t_d}$ and the discrete value $\tau_d$ of the remaining detectors relative to the reference detector should also be within an SNR time series precision $\Delta_{\rho} / 2$. When $\texttt{PyCBC}$ actually calculates this mapping, it will randomly draw $10^7$ samples (this is a user chosen number) from the isotropic priors of $\alpha$ and $\delta$, and take the mean of the $tc$'s prior as the time of signal arrival in the mapping calculation. In this way, we can capture the details of SNR time series. Since the accuracy of the discretized time delay is limited, the corresponding sky position is also a discrete area (see Fig.~1 of \citep{Roulet:2024hwz}). The prior of the discrete sky area corresponding to each discretized time delay $\tau_d$ is

\begin{subequations}
\label{eq:group_prior_tau}
\begin{align}
\Pi(\tau) &= \int_{D(\tau)} \pi(t_c, \hat{\mathbf{k}}^{\text{GEO}}) \, dt_c \, d\hat{\mathbf{k}}^{\text{GEO}} \label{eq:pi_integral} \\
          &= \int_{D(\tau)} \pi(t_c) \cdot \pi(\hat{\mathbf{k}}^{\text{GEO}}) \, dt_c \, d\hat{\mathbf{k}}^{\text{GEO}} \label{eq:factorized_prior} \\
          &\approx \frac{N_{\text{samples in } D(\tau)}}{N_{\text{samples of sky}}}.\label{eq:t_sky_prior}
\end{align}
\end{subequations}

For the proposal distribution of signal arrival time, when its prior is a uniform distribution, we can directly approximate it using the squared amplitude of the SNR time series,

\begin{equation}
q(\theta_{\text{ext}}^{i}) = q(t_{c}^{i}, \alpha^{i}, \delta^{i})
\propto \exp\left( 
    \frac{1}{2} \sum_d 
    \bigl| \rho_d(t_c^i, \hat{\mathbf{k}}_i^{\text{GEO}}) \bigr|^2 
\right). \label{eq:t_proposal}
\end{equation}

According to the definition of importance sampling, the importance weight of ($t_{c}$, $\alpha$, $\delta$) is the ratio of the target distribution, which is the conditional posterior of ($t_{c}$, $\alpha$, $\delta$), to the proposal distribution,

\begin{equation}
w_i \propto \frac{\Pi(\tau_i) \cdot \overline{\mathcal{L}}_{\phi D_L}}{q(t_{c}^{i}, \alpha^{i}, \delta^{i})}, \label{eq:t_sky_weights}
\end{equation}
now we can marginalize over the extrinsic parameters except for the inclination and polarization angles.

During the marginalization of multiband likelihood, we carry out the methods described in Sec.~\ref{subsec:multiband_margin} for 3G detectors. Then we add the LISA likelihood contribution for each marginalization sample. For phase marginalization, we use a common reference frequency for both LISA and 3G detectors (5 Hz). For marginalization of arrival time and localization, we can use ($t_{c}$, $\alpha$, $\delta$) corresponding to maximum SNR in 3G samples for all the importance samples to accelerate.

\subsection{\label{subsec:test_specific}Test with A Specific Multiband Signal}

\begin{figure*}
    \centering
    \vspace*{-0.5cm}
    \includegraphics[width=1.0\textwidth]{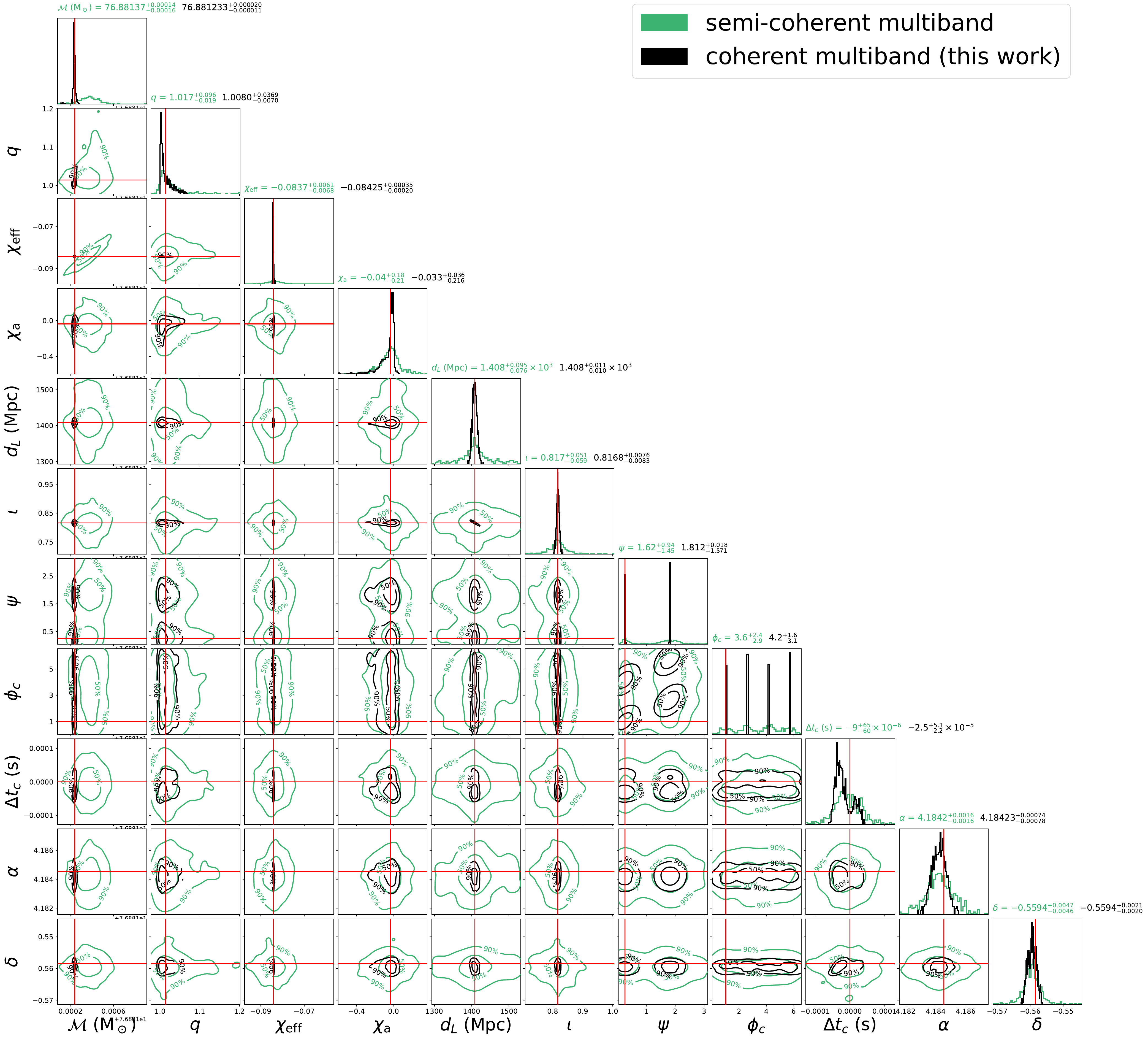}
    \caption{The comparison between coherent and semi-coherent multiband parameter estimation for the same signal from $\texttt{GWTC-3}$ population, $T_{obs}=1$ year for LISA. The semi-coherent one is using the 3G detector’s posterior as LISA’s prior. The coherent one is our new method. The introduction of non-physical degrees of freedom arises from the `3G detector posterior as LISA prior' procedure. Specifically, systematic biases introduced during this process - such as those from kernel density estimation (KDE) used to approximate the 3G posterior - can lead to suboptimal multiband parameter estimation. This limitation stems from the inherent approximation errors in propagating 3G posterior distributions to LISA. Our new coherent method doesn’t have this extra step, so we can avoid any non-physical degree of freedom.}
    \label{fig:comparision_kde_coherent}
\end{figure*}

\begin{figure*}
    \centering
    \vspace*{-0.5cm}
    \includegraphics[width=1.0\textwidth]{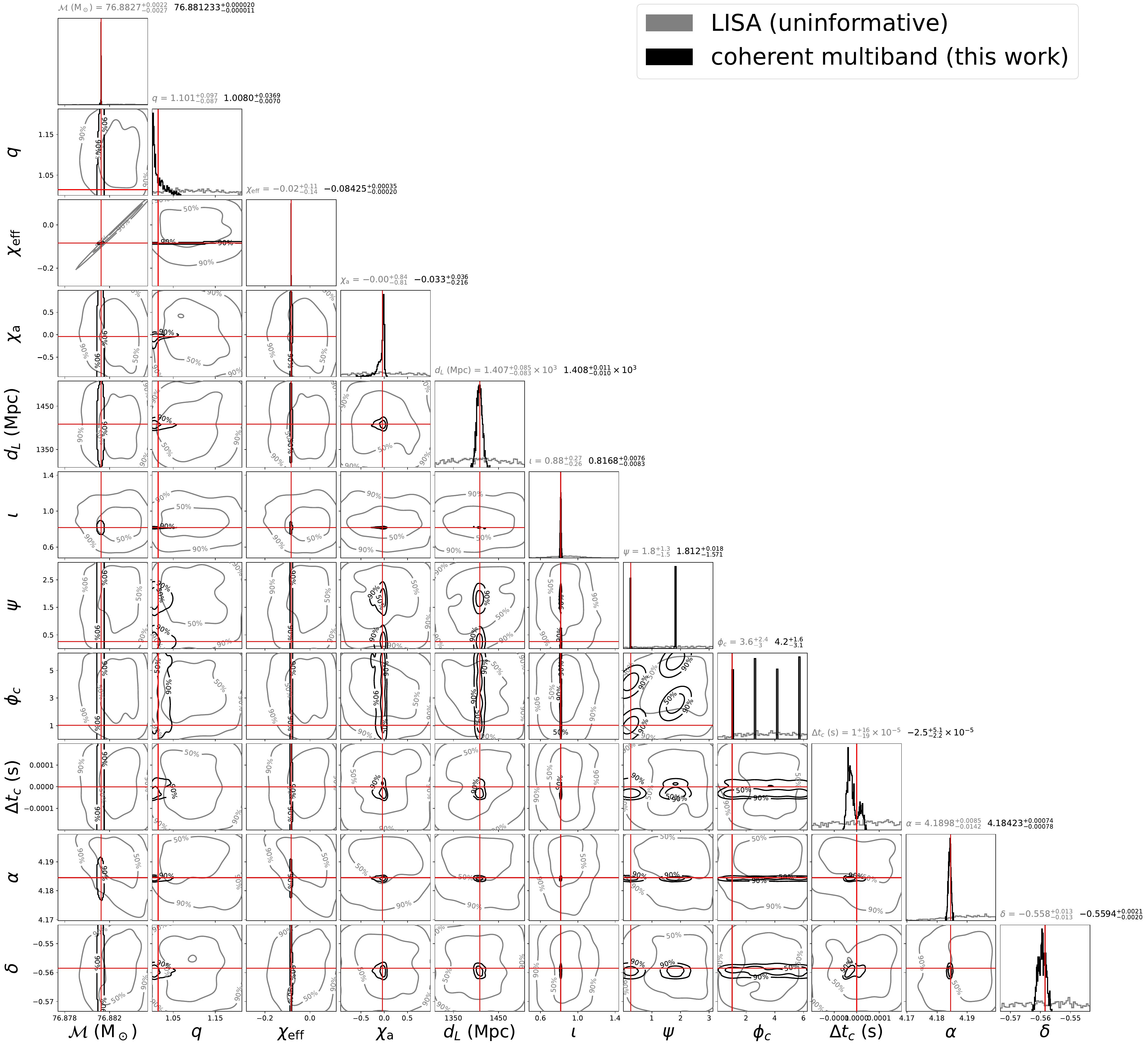}
    \caption{The comparison between coherent multiband parameter estimation and LISA-only parameter estimation for the same signal in Fig.~\ref{fig:comparision_kde_coherent}, also $T_{obs}=1$ year for LISA. The coherent one is our new method. Due to the low-SNR and so many local maxima on LISA's likelihood surface, just using LISA data and no information from ground-based detectors, makes it almost impossible to get useful information in LISA-only case.}
    \label{fig:comparision_blind_coherent}
\end{figure*}

Most of the previous multiband parameter estimation papers are based on FIM analysis \citep{Liu:2020nwz,Gupta:2020lxa,Zhao:2023ilw,Goncharov:2023woe,Seymour:2022teq,Baker:2022eiz,Muttoni:2021veo,Wu:2024yno,Tahelyani:2024cvk,DuttaRoy:2025gnu}. Since the FIM method relies on linear signal approximations, it essentially performs a low-order Taylor expansion of the likelihood function around the true parameters and assumes that a multidimensional Gaussian distribution can well represent the posterior. These prerequisites make FIM only applicable to high-SNR signals \citep{Vallisneri:2007ev}. Moreover, FIM-based multiband parameter estimation separately calculates the Fisher matrices for LISA and ground-based GW detectors and then directly adds their covariance matrices. In the later section, the SNR calculations in Fig.~\ref{fig:optimal_snr_lisa_xg} show that LISA+3G multiband observations will face extremely low-SNR LISA signals. Such low-SNR LISA signals are not suitable for FIM analysis, leading to deviations in the final multiband FIM results.

In recent years, some multiband articles based on Bayesian parameter estimation have appeared \cite{Toubiana:2022vpp,Klein:2022rbf}, analyzing a small number of signals drawn from sBBH populations. As mentioned in Sec.~\ref{sec:intro}, they first perform Bayesian parameter estimation on ground-based GW detector data and then use its posterior as the prior for LISA. Existing methods exclude extrinsic parameters from the ground-based detector's full posterior distribution and neglect coordinate transformations between 3G and LISA detectors. The challenge of accurately representing the 3G posterior as LISA's prior is further complicated by the complex extrinsic parameter distributions, even at high SNR, which can exhibit multi-modal or non-Gaussian shapes difficult to capture with standard parametrization. Such omissions introduce prior biases that disproportionately affect low-SNR signal inferences, leading to systematic uncertainties in posterior estimation and astrophysical interpretation \citep{Vitale:2017cfs,Toubiana:2020cqv}.

\begin{table}[htbp] 
    \caption{Prior distributions for coherent multiband, 3G-only, and LISA-only parameter estimation. The central values $\mathcal{M}, q, \chi_{\text{eff}}, t_c, \alpha, \delta, D_L$ in the last column are the true injected values. For the semi-coherent multiband parameter estimation, the 3G-only parameter estimation in the first step is also using the prior in this table. We use wide enough prior for almost all parameters (not just nearby true values), the narrower prior for ($t_c, \alpha, \delta$) is to save the computational time, those parameters are well measured by 3G detectors. We've made sure all the multiband runs will not hit the boundaries of prior.} 
    \label{table:prior_range}
    \begingroup
    \footnotesize     
    \setlength{\tabcolsep}{3pt} 
    % \noindent 
    \begin{tabularx}{\columnwidth}{|c|c|>{\centering\arraybackslash}X|}
    \hline
    \textbf{Name} & \textbf{Prior Type} & \textbf{Range} \\
    \hline
    $\mathcal{M}$    & uniform        & $[\mathcal{M} \pm 0.1]$ \\ 
    \hline
    $q$              & uniform        & $\begin{cases} [q \pm 0.2] & \text{if } q \!\ge\! 1.2 \\ [1, q \!+\! 0.2] & \text{else} \end{cases}$ \\ 
    \hline
    $\chi_\text{eff}$ & uniform        & \makecell{$[\max(\chi_\text{eff} \!-\! 0.2, -1),$\\ $\min(\chi_\text{eff} \!+\! 0.2, 1)]$} \\ 
    \hline
    $\chi_\text{a}$  & uniform        & $[-1, 1]$ \\
    \hline
    $t_c$ (s)        & uniform        & $[t_c \pm 0.0002]$ \\ 
    \hline
    $\alpha$         & uniform angle & $[\alpha \pm 0.015]$ \\ 
    \hline
    $\delta$         & cos angle     & $[\delta \pm 0.015]$ \\ 
    \hline
    $\psi$           & uniform angle & $[0, \pi)$ \\
    \hline
    $D_L$ (Mpc)      & uniform radius& $\begin{cases} [D_L \pm 100] & \text{if } D_L \!\ge\! 100.001 \\ [0.001, D_L \!+\! 100] & \text{else} \end{cases}$ \\ 
    \hline
    $\iota$          & sin angle     & $[0, \pi]$ \\
    \hline
    $\phi_c$         & uniform angle & $[0, 2\pi)$ \\
    \hline 
    \end{tabularx} 
    \endgroup 
\end{table}

We compare our coherent multiband method with ``3G posterior as LISA prior'' method (semi-coherent method, there will be the non-physical degree of freedom when representing 3G posterior as LISA prior) on a specific multiband signal from Sec.~{\ref{sec:population_runs}}, which LISA SNR is around 5 and $T_{obs}$=1 year for LISA, using the prior defined in Table.~\ref{table:prior_range}. As can be seen from the posterior comparison in Fig.~\ref{fig:comparision_kde_coherent}, the coherent multiband method proposed in this paper can achieve a more robust and sensitive multiband Bayesian parameter estimation (e.g. narrower posterior). The continuous wave search in ground-based detectors \citep{Wette:2023dom} already showed that coherent analysis (which can maintain the consistency of amplitude and phase for the whole signal template and data in all data segments) is much more sensitive. Multiband is a special case in which we can treat the LISA and 3G data as two data segments for the same GW signal. Actually, previous Bayesian multiband papers even used a simpler method to represent 3G posterior, they used a Gaussian prior with covariance matrix computed from the ground-based detectors' samples for a subset of parameters \cite{Toubiana:2022vpp}. We used KDE to capture the full covariance matrix and relax the Gaussian assumption for the semi-coherent multiband run as a comparison to our coherent method.

Compared to the LISA-only results, which use an uninformative prior rather than a 3G-informed prior, we can further recognize the reliability of the coherent method. In Fig.~\ref{fig:comparision_blind_coherent}, due to the low SNR of LISA and the complex likelihood, the LISA-only analysis yields minimal informative constraints, as the posterior distributions remain almost statistically indistinguishable from the prior assumptions.

\section{\label{sec:population_runs}Coherent multiband analysis of stellar-mass BBH population}

In this section, we further evaluate the performance of our coherent multiband method using a realistic population model for sBBH. First, we outline the population model assumptions (\ref{sec:population_model}) and the multiband definition (\ref{sec:mock_data}) adopted in this work. Subsequently, we present key statistical results derived from this model, including the SNR distribution (\ref{subsubsec:snr_dist_3g_lisa}), multiband detection rate (\ref{subsubsec:det_rate_multiband}), and the distribution of merger timescales (\ref{subsubsec:t_merge_multiband}). We then investigate the impact of the LISA mission duration on the multiband posterior inference (\ref{subsubsec:t_obs_effect_multiband}) and assess the robustness of our method when applied to extremely low-SNR LISA sBBH signals (\ref{subsubsec:low_snr_multiband}). Finally, we demonstrate insights that can be extracted through population-scale Bayesian multiband analysis (\ref{subsubsec:pop_scale_multiband}). In this section, we make several observation scenarios for the duration of the LISA mission and randomly draw sBBHs from the \texttt{GWTC-3} population with a low SNR threshold.

All multiband runs in this work employ the \texttt{PyCBC Inference} framework \citep{Biwer:2018osg} with the \texttt{nessai} sampler \citep{nessai,Williams:2021qyt,Williams:2023ppp} to explore the parameter space of sBBHs. \texttt{nessai} is a nested sampling algorithm that integrates normalizing flows to enhance sampling efficiency, specifically designed for computationally intensive problems where the likelihood evaluation is expensive. Each coherent multiband analysis typically requires a few days of computation on 8 CPU threads.

\subsection{\label{sec:population_model}The population models}

We briefly describe the population model that we are using to generate our mock dataset; more details can be found in \cite{Wu:2022pyg}. This paper focuses on the analysis of non-precessing, quasi-circular, and stellar-mass binary black holes (sBBHs). These sBBHs can be fully characterized by 11 parameters, including intrinsic parameters (such as the mass of each black hole $m_{1,2}$ and their spin parallel to the orbital angular momentum $\chi_{1z,2z}$) and extrinsic parameters, such as the coalescence time $t_{c}$, coalescence phase $\phi_c$, sky localization ($\alpha, \delta$), luminosity distance $D_{L}$, polarization angle $\psi$ and inclination angle $\iota$. Precession is not included in this analysis, as the package \texttt{BBHx} \citep{Katz:2020hku} used to generate the LISA signal does not support precession (up to now, precession is an open question for LISA data analysis). This package is based on the \texttt{IMRPhenomD} and \texttt{IMRPhenomHM} models and accounts for the LISA detector response and time-delay interferometry, we add TDI-2 support for it based on Eq.~(33) in \citep{Babak:2021mhe}. For intrinsic parameters, mass and mass ratio follow a \textit{ power law + peak} model (as in Fig.~10 of \citep{KAGRA:2021duu}), while spin amplitudes use the distribution from Fig.~15 of the same work, projected onto the orbital angular momentum under isotropic orientation to yield aligned spins. Extrinsic parameters (position, polarization angle, inclination angle, coalescence phase) are assigned isotropic/uniform distributions. Following \cite{Wu:2022pyg}, we compute the sBBH redshift and luminosity distance distribution via the convolution of the star formation rate (SFR) and time-delay distributions to evaluate 3G detector sensitivity to high-redshift CBC signals,

\begin{equation}
\begin{aligned}
\dot{\rho}(z)=\dot{\rho}_{0} f(z) & \propto \int_{\tau_{\min }}^{\infty} \dot{\rho}_{*}\left[z_{f}(z,\tau)\right] P(\tau) d \tau \\
& \propto \int_z^{\infty} \dot{\rho}_{*}\left(z_{f}\right) P\left[\tau\left(z,z_{f}\right)\right] \frac{d t\left(z_{f}\right)}{d z_{f}} d z_{f}, \label{eq:merger_rate_den}
\end{aligned}
\end{equation}
here $\dot{\rho}_{0}$ is the local coalescence rate density (in the unit of $\text {Mpc}^{-3} \mathrm{yr}^{-1}$), and $f(z)$ is the normalized coalescence rate density. The SFR $\dot{\rho}_{*}$ is in the unit of $M_{\odot} \mathrm{Mpc}^{-3} \mathrm{yr}^{-1}$. $P(\tau)$ is the probability distribution of the time delay $\tau$. This time delay $\tau=t(z)-t\left(z_{f}\right)$ refers to the total time from the formation of the binary progenitors (when redshift is $z_{f}$) to the merger of the compact binary due to GW emission (when redshift is $z$). This delay is determined by the difference between the lookback time of $z$ and $z_{f}$, and the lookback time at redshift $z$ is defined as
\begin{equation}
t(z)=\frac{1}{H_{0}} \int_{z}^{\infty} \frac{d z}{(1+z) \sqrt{\Omega_{\Lambda}+\Omega_{\mathrm{m}}(1+z)^{3}}}, \label{eq:lookback_time}
\end{equation}
where ${H_{0}}$ is the Hubble constant, and $\Omega_{\Lambda}$ and $\Omega_{\mathrm{m}}$ are the densities of dark energy and nonrelativistic matter, respectively. In this paper, we assume the standard $\Lambda \mathrm{CDM}$ cosmology \citep{Planck:2015fie} using ${H_{0}}=67.74 \mathrm{~km} \mathrm{~s}^{-1} \mathrm{Mpc}^{-1}$, $\Omega_{\Lambda}=0.6910$, and $\Omega_{\mathrm{m}}=0.3075$. We use the inverse delay model $P\left(\tau \right) \propto 1 / \tau$. In order to obtain the distribution of the event rate of sBBHs as a function of redshift in the detector frame, we need to multiply the coalescence rate density expressed by Eq.~(\ref{eq:merger_rate_den}) by the comoving volume element $d V(z) / d z$, and then divide it by $1+z$ caused by the time dilation, so we get the following equation
\begin{equation}
\frac{d R}{d z}=\frac{\dot{\rho}_{0} f(z)}{1+z} \frac{d V(z)}{d z}, \label{eq:merger_rate}
\end{equation}
where the comoving volume element $d V(z) / d z$ is described by
\begin{equation}
\frac{d V(z)}{d z}=\frac{c}{H_{0}} \frac{4 \pi D_{\mathrm{L}}^{2}}{(1+z)^{2} \sqrt{\Omega_{\Lambda}+\Omega_{\mathrm{m}}(1+z)^{3}}}, \label{eq:comoving_volume}
\end{equation}
where $c$ is the speed of light in the vacuum, and $D_{\mathrm{L}}$ is the luminosity distance between the sBBH and the detector, which is defined as
\begin{equation}
D_L=(1+z) \frac{c}{H_{0}} \int_{0}^{z} \frac{d z}{\sqrt{\Omega_{\Lambda}+\Omega_{\mathrm{m}}(1+z)^{3}}}. \label{eq:luminosity_distance}
\end{equation}

For the local merger rate of sBBHs, we choose the rate based on LVK's population paper of \texttt{GWTC-3} \citep{KAGRA:2021duu} and the public presentation~\footnote{Webinar: The population of CBC inferred using GWs through \texttt{GWTC-3} \url{https://dcc.ligo.org/LIGO-G2102458/public}}. We choose 22 Gpc$^{-3}$yr$^{-1}$ as median local merger rate. Then, we draw the redshift of the GW signal from this distribution, with an upper bound of $z_{\max}=20$.

We can get the average time interval between two nearby GW signals from sBBHs as,
\begin{equation}
\overline{\Delta t}=\left[\int_{0}^{z_{\max }} \frac{d R}{d z}(z) \mathrm{d} z\right]^{-1}. \label{eq:average_time_interval}
\end{equation}

Assuming the median local merger rate, we find the average time interval for sBBHs is around 359.4 s. We use this value to determine how many signals to simulate in a given time. After determining the number of sBBH signals, the coalescence time is selected uniformly and randomly within the given time.

\subsection{\label{sec:mock_data}Mock data generation}

With the population model in Sec.~\ref{sec:population_model}, now we can simulate the corresponding data from LISA and 3G ground-based detectors. In this paper, we consider the ideal situation, where ground-based and space-borne GW detectors only contain detector noise plus a GW signal, without considering the situation where many signals overlap in actual data, which is considered in the LISA Global Fit \citep{Cornish:2005qw,Robson:2017ayy,Littenberg:2020bxy,Littenberg:2023xpl,Rosati:2024lcs,Strub:2024kbe,Katz:2024oqg,Deng:2025wgk}. However, when generating LISA data, we used PSD with a DWD confusion noise contribution (as shown in Fig.~\ref{fig:multiband_asd}), which can simulate the loss in SNR to some extent. Although most sBBHs do not pass through the region affected by DWD confusion noise due to their lighter masses compared to supermassive or intermediate-mass black hole binaries. In addition, we have also corrected the relative position of the LISA detector and the Earth in the SSB frame in \texttt{BBHx}, so that LISA basically follows at a position 20° behind the Earth within the time range of data simulation.

In the LISA noise simulation, we selected approximately orthogonal TDI-2's A, E, and T channels \citep{Otto:2015erp,Tinto:2020fcc}, and for ground-based detectors, we selected a detector network of one Einstein Telescope (ET) and two Cosmic Explorer (CE). The ET adopts the triangular scheme \citep{Hild:2009ns} with a 10 km arm length, and the CE adopts the 40km+20km scheme \citep{Evans:2023euw}.

To consider the impact of the duration of the LISA mission on the estimation of multiband parameters, we considered six different observation scenarios ($T_{obs}$=1, 2, 3, 3.75, 4.5, 7.5 years). $T_{obs}$=1 yr datasets are to investigate whether multiband analysis can be achieved in the first year of LISA data analysis \citep{Seoane:2021kkk}; $T_{obs}$=2 yrs datasets are to mimic the length of the latest \texttt{LDC-Yorsh} dataset~\footnote{LISA Data Challenge 1b: Yorsh \url{https://lisa-ldc.lal.in2p3.fr/challenge1b}}; $T_{obs}$=3, 3.75, 4.5 yrs datasets are set according to Table 4 in \citep{Seoane:2021kkk}, which are 4, 5, and 6 years multiplied by a duty cycle of 0.75; $T_{obs}$=7.5 yrs datasets are the maximum LISA possible lifetime multiplied by that duty cycle of 0.75.

In this paper, the definition of multiband is that if the signals can be detected by ground-based detectors within the time when LISA starts observing, and the LISA observation ends plus an additional $T_{wait}$=5 yrs (so $T_{obs} + T_{wait}$ for ground-based detectors to wait for the final merger signal of sBBHs), also the SNR in LISA band $\rho_{LISA}$ should be higher than 3. To account for potential variability in the observed population sampling, we generated 20 distinct datasets per $T_{obs}$ scenario using separate random seeds, resulting in a total of 120 datasets.

\subsection{\label{sec:population_results}Results of population run}

\subsubsection{\label{subsubsec:snr_dist_3g_lisa}SNR Distribution of Multiband in 3G and LISA}

Based on the Sec.~\ref{sec:population_model} population model and the Sec.~\ref{sec:mock_data} detector network settings and observation scenarios, we calculate whether each randomly simulated signal satisfies our definition (see the last paragraph in Sec.~\ref{sec:mock_data}) of multiband signals. Since the number of events to be simulated is too large ($T_{obs}$ in years divided by $\overline{\Delta t}$ in minutes, see Eq.~(\ref{eq:average_time_interval})), strictly calculating the LISA TDI response and optimal SNR for each signal would result in excessive calculations. We adopt a more efficient but still accurate method. We first calculate the horizon distance of LISA for sBBHs of equal mass in advance based on the sky-averaged sensitivity curve and signals for LISA \citep{Robson:2018ifk}, that is, the farthest distance that LISA can see sBBHs of this mass given an SNR threshold ($\rho_{LISA}$=3 for us) and observation time. Each time we randomly draw a signal from the population model, we find the corresponding horizon distance using twice the maximum source-frame component mass as the upper limit of the total mass. If the luminosity distance $D_L$ of the simulated event is greater than this horizon distance, or the starting frequency of the signal is higher than LISA's high-frequency cutoff, or the time required to wait until the merge $T_{merge}$ exceeds 5 years, then the signal is skipped. If the conditions are met, the LISA TDI detector response and optimal SNR are calculated based on the complete signal parameters. If the optimal SNR is still higher than the given threshold, multiband and 3G-only parameter estimation will be launched for the signal. The entire process is fully automated by \texttt{PyCBC} workflow scripts.

We summarize the SNR distribution of multiband events in our datasets for LISA and the ET+2CE GW detector network in Fig.~\ref{fig:optimal_snr_lisa_xg}. The SNR in LISA measures the strength of the signal during the very early inspiral stage, while the SNR in the ET+2CE network measures the strength of the same signal during the late inspiral and merge stages. The scatter plot only displays the multiband signal with $T_{obs}$=7.5 years. The color of the plot indicates the time elapsed between the signal's observation by LISA and its merger in 3G detectors. It can be seen that most multiband signals will merge in about 10 years, which we will analyze in more detail later.

The one-dimensional marginal distributions in the upper and right parts are obtained by summing the results of all random number seed simulations, which can avoid the statistical deviation caused by the small number of signals in a certain dataset. The different lines on the right correspond to different LISA observation durations $T_{obs}$. It can be seen that $\rho_{LISA}$ is concentrated around 4, and in most cases it will be lower than 10, and it is not sensitive to different LISA observation durations. In contrast, the SNR in ground-based detectors will be very high, because their sensitivity is an order of magnitude higher than that of the current Advanced LIGO. Most signals are concentrated at about 600 for $\rho_{ET+2CE}$, with a tail extending to 3000. This is in stark contrast to one frequency band with a very high SNR and another with an extremely low SNR. This poses a great challenge for multiband signal data analysis. Moreover, previous multiband analysis method is only applicable when $\rho_{LISA}$ is greater than 5 \citep{Toubiana:2022vpp}, and according to the results of this figure, a large part of the signal is not applicable.

\begin{figure}[h]
	\includegraphics[width=0.5\textwidth]{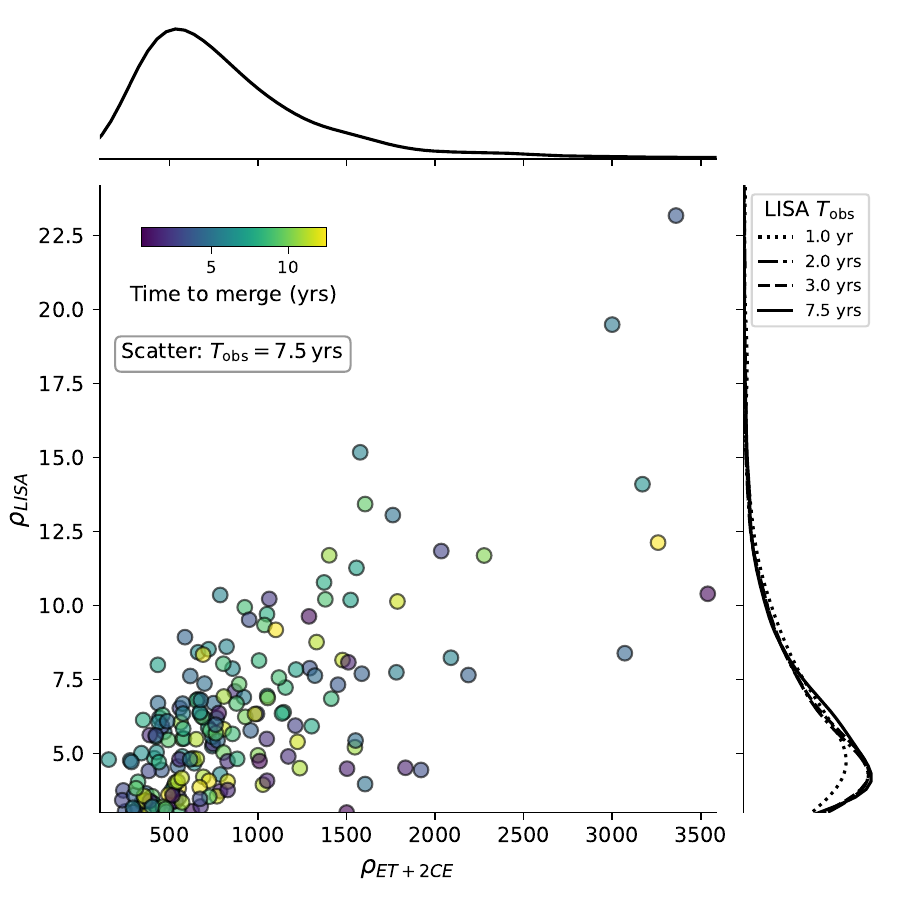}
    \caption{This figure presents the optimal SNR distribution of multiband stellar-mass BBHs (sBBHs) in LISA and the next-generation ground-based detector network, as well as the time required for coalescence. The color of the scatter points represents the time required for coalescence. The grey lines in marginal distributions represent the probability density functions (PDF) corresponding to the scatter points. It can be observed that within the LISA frequency band, the signal-to-noise ratio (SNR) is predominantly below 8, with the majority exhibiting an SNR below 5. The SNR of these signals in the next-generation ground-based detector network is primarily concentrated around 600. From the color distribution, it can be seen that there is no clear trend between the required merging time and the corresponding SNR. We set 3 as the minimum SNR cuto-ff for LISA.}
    \label{fig:optimal_snr_lisa_xg}
\end{figure}

\subsubsection{\label{subsubsec:det_rate_multiband}Detection Rate of Multiband Signal}

To estimate the population size detectable by multiband, we also calculated the detection rate for multiband signals in this study. The difference is that we took two LISA observation durations ($T_{obs}$=3.0 and 7.5 yrs) and three different LISA SNR thresholds ($\rho_{th}$=3, 5, 8). In Fig.~\ref{fig:event_rate_lisa}, the internal plot with error bars summarizes the external histogram and shows the median number of events, as well as the confidence intervals 68\% and 90\%. In general, longer LISA observation times and lower SNR thresholds result in more multiband signals. This is because, as can be seen from Fig.~\ref{fig:optimal_snr_lisa_xg} and the previous discussion, the bottleneck in the multiband event rate is the low SNR in the LISA band. We have also tried to reproduce the Fig.~1 in \citep{Buscicchio:2024asl}, using the results of our own simulations, and they agree statistically with each other. Because a large number of multiband sBBHs signals have a SNR of less than 5 in the LISA band, the number of events is very sensitive to low SNR thresholds. If there is some kind of data analysis method that can reduce the SNR threshold to 3 instead of 5, we will get a doubling of the signal number.

\begin{figure}[h]
	\includegraphics[width=0.53\textwidth]{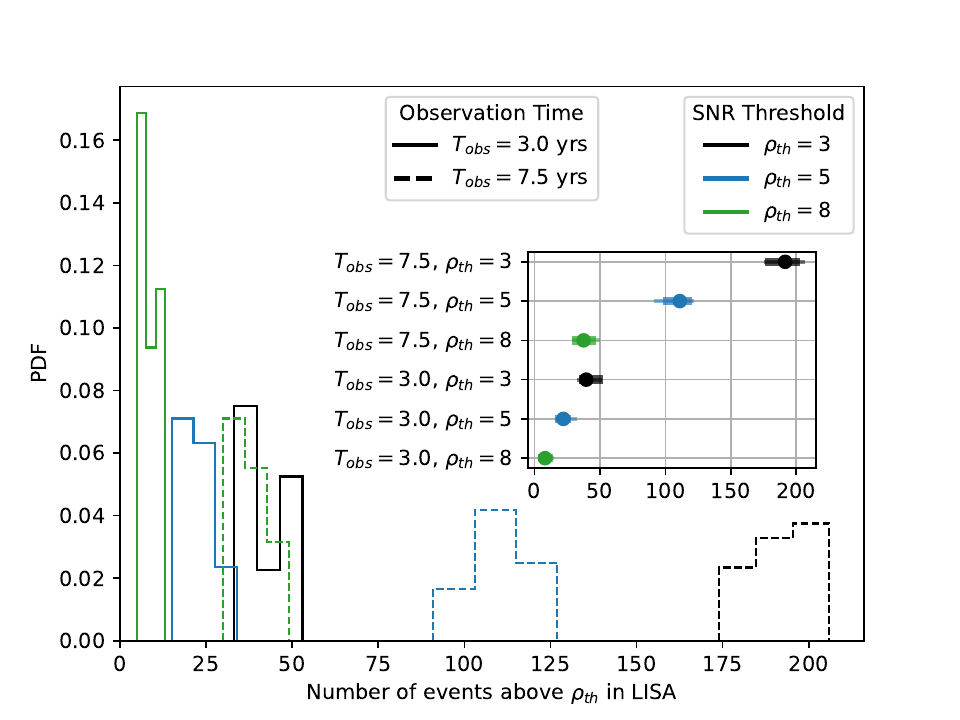}
    \caption{This figure illustrates the multiband event rates for different LISA observation durations and SNR thresholds. The results are based on population simulations from 20 different random seeds. Only two LISA mission duration scenarios (4 years and 10 years) are plotted, assuming the LISA duty cycle is 0.75, resulting in LISA observation times of 3 years and 7.5 years, respectively. As illustrated in Fig.~\ref{fig:optimal_snr_lisa_xg}, the distribution of SNR reveals that sBBHs are concentrated below 8 in the LISA band. This suggests that the limitation of the multiband event rate lies within LISA rather than in 3G detectors. In general, an increase in the length of the LISA observation time and a decrease in the SNR threshold result in a higher multiband event rate. If a conservative SNR threshold of 8 and a LISA fiducial mission duration of 4 years are selected, the event rate is less than 10. Given that approximately half of the sBBH events have an SNR below 5 in the LISA band, the event rate is contingent upon the lowest SNR threshold that can be attained through data analysis techniques. The most state-of-the-art multiband method can only analyze events with a LISA SNR above 5. Consequently, if there were a more sensitive method that could reduce the threshold to 3, the number of available multiband events would nearly double.}
    \label{fig:event_rate_lisa}
\end{figure}

\subsubsection{\label{subsubsec:t_merge_multiband}The Time to Merger for Multiband Signals}

For multiband GW events, another concern is how long this signal needs to merge. On the one hand, this can be used as a guide for subsequent observations by ground-based GW detectors, such as the time should be online as far as possible to avoid missing valuable multiband signals; on the other hand, ground-based detectors and data analysts cannot endure prolonged waiting periods.

Fig.~\ref{fig:merge_time_dist} summarizes the required merging time $T_{merge}$ (the time required for multiband signals from the beginning of LISA observations to the merge in ground-based detectors’ frequency band), and we consider the effect of different LISA observation durations in the datasets. It can be seen that as $T_{obs}$ increases, the distribution of the waiting time $T_{merge}$ as a whole shifts to longer times. This is expected, because signals that still pass the SNR threshold with shorter LISA observation durations must have higher amplitudes, and sBBHs generally have higher amplitudes in the part that is closer to the merger, so they need to wait for a shorter time. This can also be seen in Fig.~\ref{fig:multiband_asd}, where the two signals that merge fastest are both in the very high frequency part of LISA. Overall, the majority of multiband signals require a merger time of about 5 years, matches the peak of Fig.~2 in \citep{Buscicchio:2024asl}, the key distinction is that they have many events need $10^{2}$ to $10^{3}$ years to merge, because they didn't set a $T_{merge}$ requirement for ground-based detectors.

\begin{figure}[h]
	\includegraphics[width=0.53\textwidth]{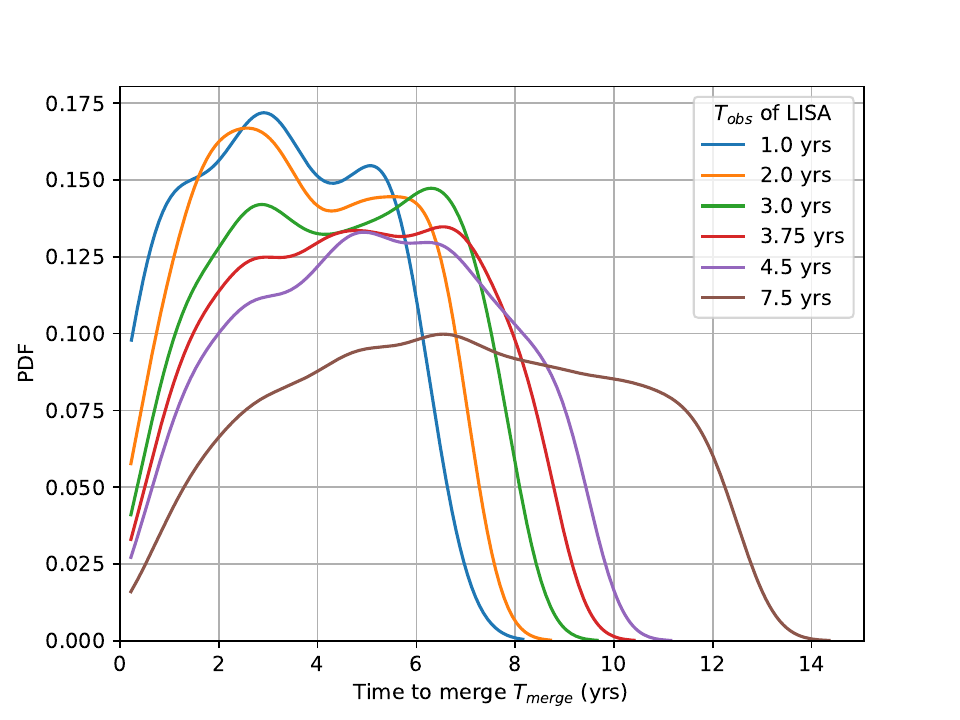}
    \caption{In this figure, we show the probability distribution of the time required for multiband signals from the beginning of LISA observations to the merge in ground-based detectors’ frequency band, under different LISA observation durations. Overall, the multiband signals considered in this work ($\rho_{LISA} \ge 3$ and merged before the end of LISA observation plus 5 years’ waiting time) tend to merge in about 5 years. As the LISA observation duration $T_{obs}$ increases, $T_{merge}$ as a whole tends to increase over time because signals that can pass LISA's SNR threshold with shorter $T_{obs}$ will be at higher frequencies (these sBBHs signals have larger amplitudes at higher frequencies), and therefore closer to the final merger. Note that tails of those distributions are smoothed by kernel density estimation and clipped at the lower end using the minimum $T_{merge}$ in the simulation.}
    \label{fig:merge_time_dist}
\end{figure}

\subsubsection{\label{subsubsec:t_obs_effect_multiband}How Does LISA Observation Duration affect Multiband Signals?}

\begin{figure}[h]
	\includegraphics[width=0.49\textwidth]{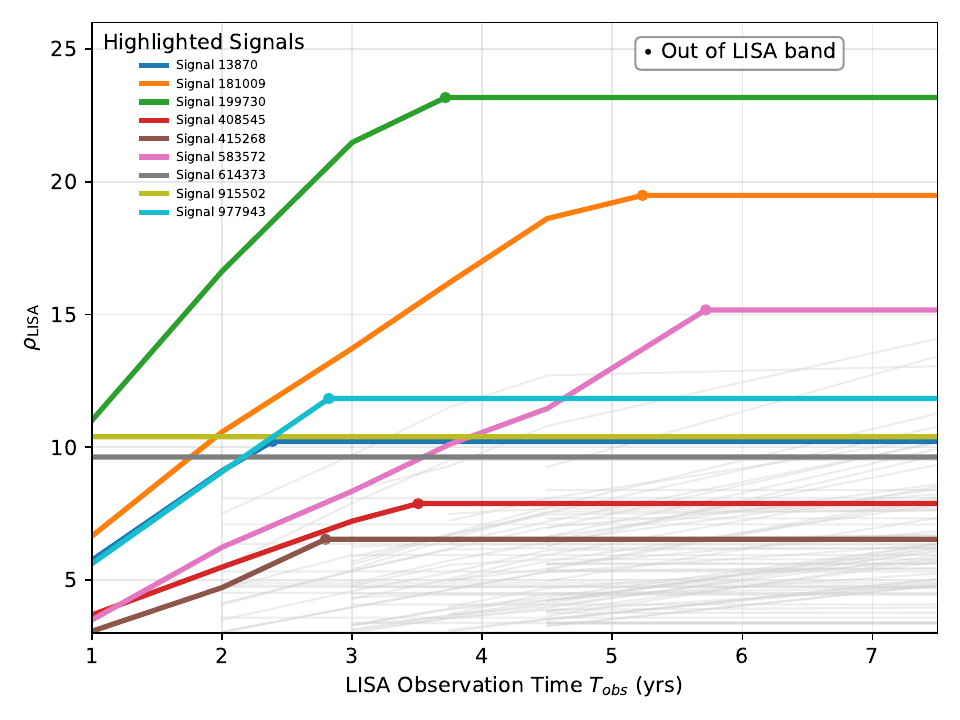}
    \caption{This figure shows the accumulation of the LISA SNR with the LISA observation time $T_{obs}$ for a set of simulated multiband observations. The colored lines mark signals with $\rho_{LISA} \ge 3$ from $T_{obs}=1$ yr dataset, and the colors of the signals exactly correspond to those in Fig.~\ref{fig:multiband_asd}. As time increases, the values of $\rho_{LISA}$ are roughly proportional to the square root of $T_{obs}$. Once a signal evolves outside of the LISA band, $\rho_{LISA}$ remains constant because these signals are no longer present in subsequent data. Two of these signals evolve out of the LISA band during the first year of LISA observation. The gray lines are the other multiband signals that appear in the subsequent $T_{obs}$ datasets. Only points at $T_{obs}$ in \{1.0, 2.0, 3.0, 3.75, 4.5, 7.5\} years are calculated in this figure.}
    \label{fig:lisa_snr_evolution}
\end{figure}

Here we investigate the growth of multiband signal SNR in the LISA band with observation time, a factor previously unexplored in prior multiband studies. This analysis reveals how prolonged observations can enhance signal detectability, offering new insights for mission planning. Fig.~\ref{fig:lisa_snr_evolution} shows the cumulative plot of $\rho_{LISA}$ based on a subset of our datasets. The highlighted signals are those with $\rho_{LISA} \ge 3$ in the $T_{obs}$=1 yr dataset, and the colors correspond exactly to those in Fig.~\ref{fig:multiband_asd}. The gray background lines are multiband signals that appear in subsequent observation time datasets. Their overall trend is similar to that of the highlighted signals, but they lag slightly in time.

The circular nodes represent the signal leaving the LISA band. The reason why $\rho_{LISA}$ remains constant after this point is that the signal is no longer present in the subsequent data. When the signal is within the LISA band, $\rho_{LISA}$ is roughly proportional to the square root of the observation time, which is consistent with the general law of the SNR accumulation of the matched filtering.

\begin{figure*}
    \centering
    \vspace*{-0.5cm}
    \includegraphics[width=1.0\textwidth]{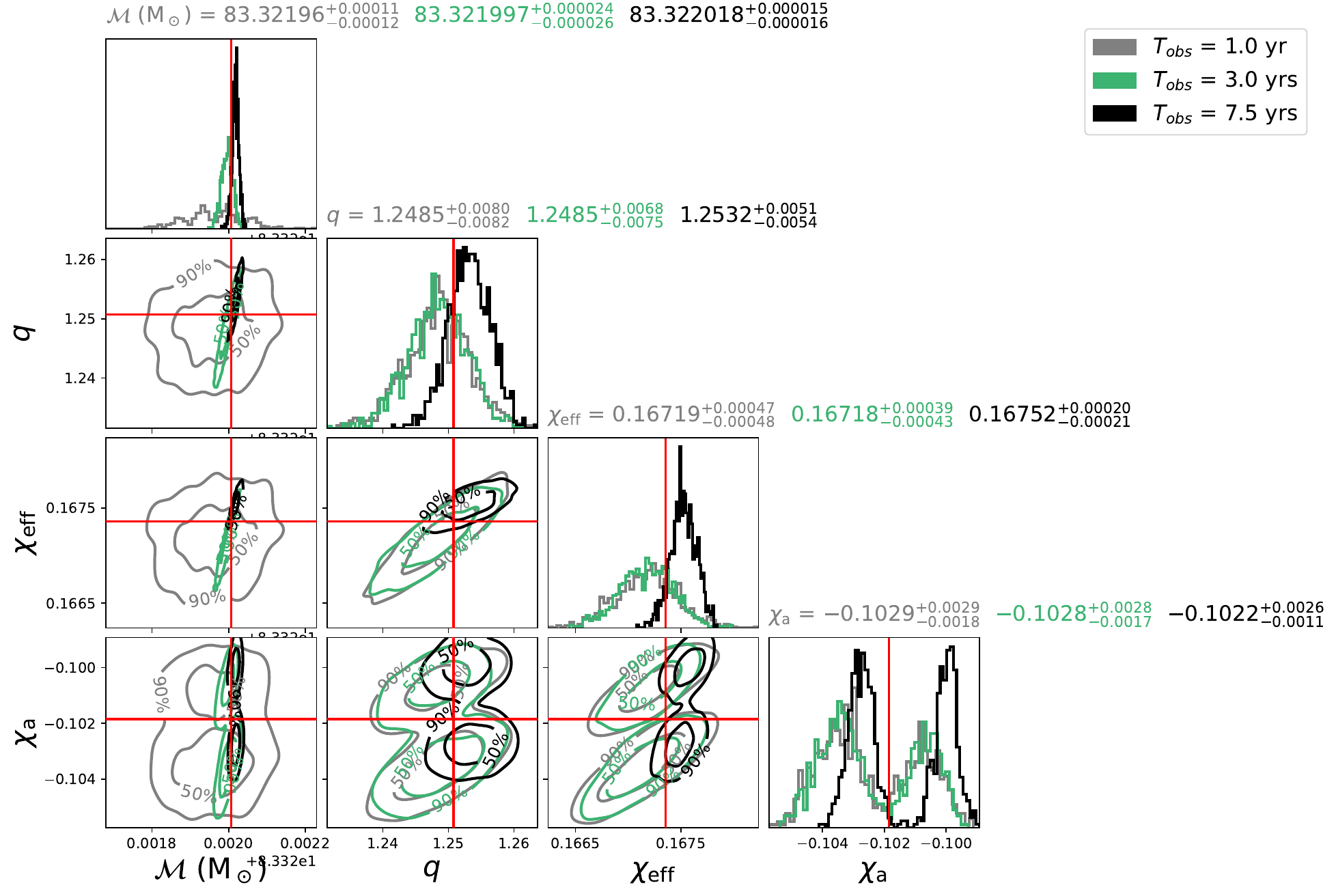}
    \caption{This figure shows the change of the multiband posterior distribution for intrinsic parameters (chirp mass $\mathcal{M}$, mass ratio $q$, effective spin $\chi_\mathrm{eff}$, asymmetric spin $\chi_\mathrm{a}$) with the LISA observation time $T_{obs}$. Here, the simulated signal numbered 583572 in the data set is used as an example; according to Fig.~\ref{fig:lisa_snr_evolution}, it can be seen that this signal stays within the LISA frequency band roughly until $T_{obs} = 6$ years. As can be seen from the figure, as $\rho_{LISA}$ increases, the posterior distribution of the intrinsic parameters becomes narrower and narrower. In particular, the chirp mass’s 90\% credible interval is improved by an order of magnitude, from $10^{-4}$ to $10^{-5}$.}
    \label{fig:multiband_different_tobs_intrinsic}
\end{figure*}

\begin{figure*}
    \centering
    \vspace*{-0.5cm}
    \includegraphics[width=1.0\textwidth]{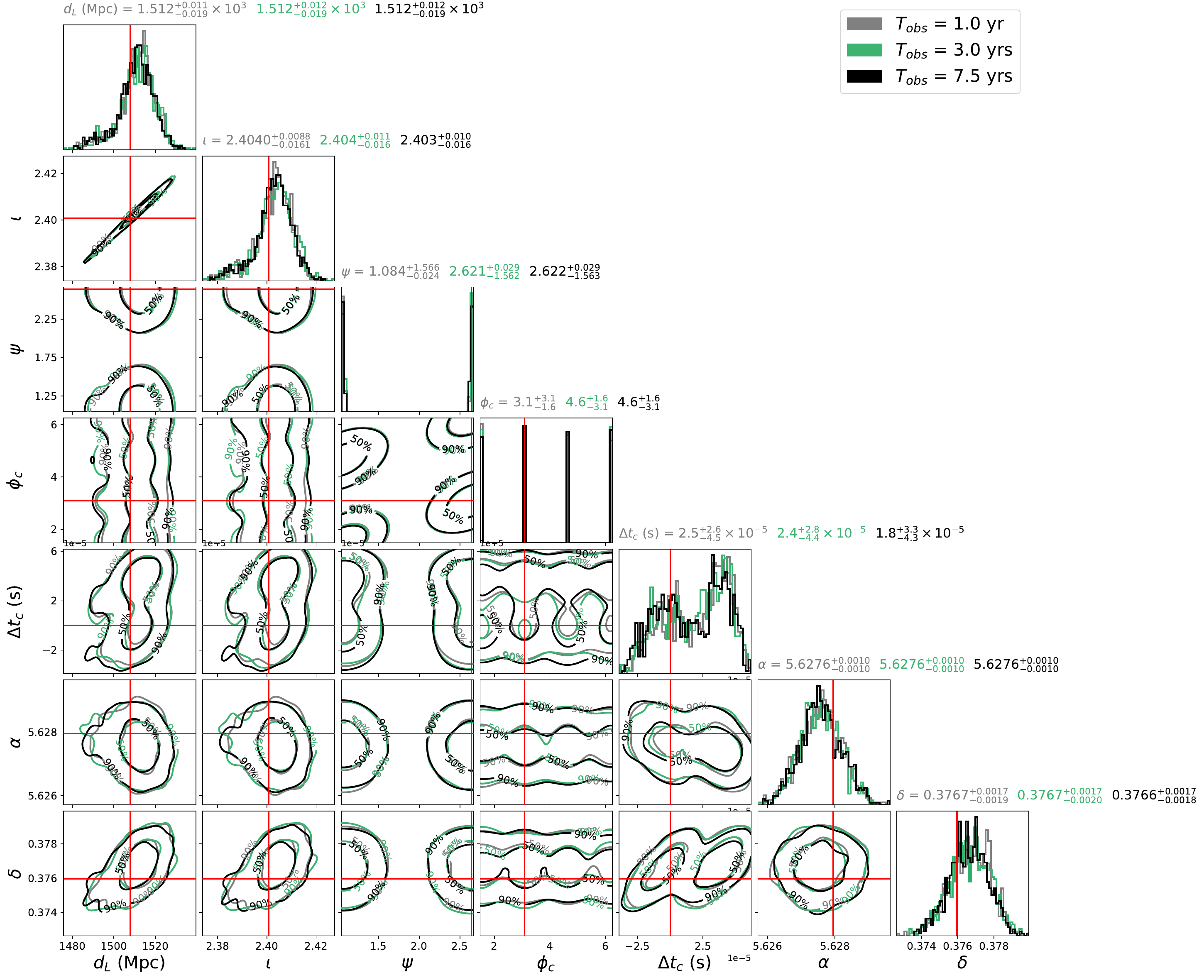}
    \caption{This figure shows the change of the multiband posterior distribution for extrinsic parameters (luminosity distance $D_L$, inclination angle $\iota$, polarization angle $\psi$, coalesence phase $\phi_c$, coalesence time offset $\Delta t_c$, right ascension $\alpha$, declination $\delta$) with the LISA observation time $T_{obs}$, for the same simulated signal in Fig.~\ref{fig:multiband_different_tobs_intrinsic}. As we can see, the posterior of extrinsic parameters for this signal are not improved by longer $T_{obs}$, because the
    ground-based detector network can constrain those parameters much better than LISA in this case.}
    \label{fig:multiband_different_tobs_extrinsic}
\end{figure*}

One interesting point is how different LISA observation durations affect the posterior distribution of multiband signals. Here, we select Signal 583572 in Fig.~\ref{fig:lisa_snr_evolution}. This signal is more suitable because it is mostly in the LISA frequency band throughout the entire observation duration, and it does not leave the LISA frequency band until about 5.6 years. Fig.~\ref{fig:multiband_different_tobs_intrinsic} and Fig.~\ref{fig:multiband_different_tobs_extrinsic} show the posterior distributions of the intrinsic and extrinsic parameters of this signal, respectively, using three different LISA observation times. It can be seen that the intrinsic parameters have all improved significantly, especially the chirp mass, which has improved by an order of magnitude. However, the extrinsic parameters have not improved because, for this signal, ground-based GW detectors can constrain these parameters better than LISA, so the additional information from LISA does not improve on this signal.

\subsubsection{\label{subsubsec:low_snr_multiband}Can We Make Use of Low-SNR Multiband Signals?}

\begin{figure*}
    \centering
    \vspace*{-0.5cm}
    \includegraphics[width=1.0\textwidth]{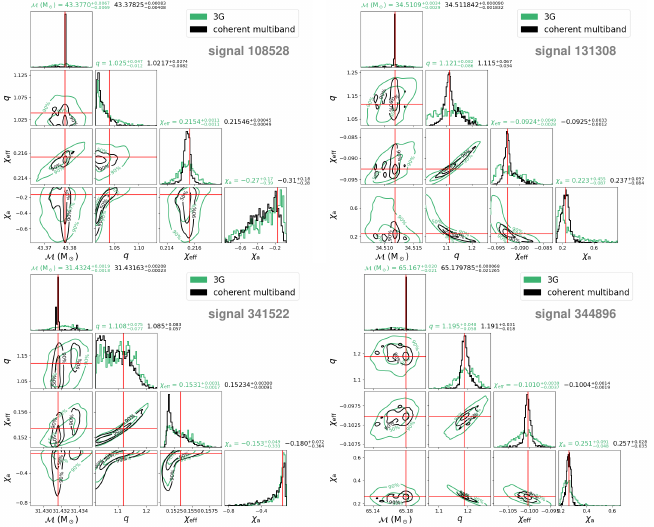}
    \caption{This figure shows the multiband and 3G-only parameter estimation results for the signal with $\rho_{LISA} \simeq 3$. It shows the intrinsic parameters. Through the new method of phase coherence and extrinsic marginalization in this paper, we have achieved information extraction from extremely low-SNR signals. Previously, the most advanced multiband method could only analyze signals with $\rho_{LISA} \ge 5$. It is worth mentioning that although a very narrow true likelihood peak can be found in chirp mass, the posterior distribution exhibits multiple small modes, because the $\rho_{LISA}$ is close to the SNR contribution of the detector noise itself (see the detector-independent Rayleigh distribution of Guassian noise's SNR from Fig.~6 in \citep{Wu:2022pyg}), so many local maxima on the likelihood surface. Other intrinsic parameters can also be improved by such low SNR signal in LISA.}
    \label{fig:xg_multiband_snr_3}
\end{figure*}

We also analyzed signals with extremely low SNR in the dataset. We randomly selected four signals with $\rho_{LISA} \simeq 3$ in Fig.~\ref{fig:xg_multiband_snr_3}. We can see that our novel method with phase coherence and marginalization over extrinsic parameters is still applicable here. Our method can accurately find the narrow true likelihood peak of the chirp mass, but it is worth mentioning that there are multiple modes around it. This is because, at such low SNR, the SNR contributed by the LISA detector noise itself is already close to it, which can cause many local maxima on the likelihood surface. From Fig.~6 in \citep{Wu:2022pyg}, we can see that the Gaussian noise's matched-filtering SNR follows a Rayleigh distribution $|\rho| \cdot e^{-|\rho|^2/2}$, which is independent of detector type, so it also applies to the LISA case (see panel (c) in Fig.~\ref{fig:strain_snr_sbbh_lisa}).

Therefore, testing the new method proposed in this paper on population-scale multiband mock data is necessary. This can reflect the actual effect of LISA+3G multiband observations as realistically as possible.

\subsubsection{\label{subsubsec:pop_scale_multiband}What Can We Learn from Population-scale Multiband Parameter Estimation?}

Fig.~\ref{fig:multiband_pop_ci_width} summarizes our population-scale multiband parameter estimation, showing the distribution of the 90\% credible interval widths of the posterior of each parameter, with the vertical axis representing the average number of events in each credible interval width bin (combining the datasets from all random seeds). We omit polarization angle $\psi$ and coalescence phase $\phi_c$ from the figure because their posteriors show multiple peaks, making the 90\% CI nearly span the entire prior (see Fig.~\ref{fig:comparision_kde_coherent} and Fig.~\ref{fig:multiband_different_tobs_extrinsic}). To compare parameter estimation precision across orders of magnitude, all parameters except mass ratio $q$, right ascension $\alpha$, and declination $\delta$ are expressed in base-10 exponential form. Colored histograms correspond to four LISA observation durations, while the gray histogram shows ET+2CE network posteriors for ground-based events from multiband signals with 7.5-year LISA observation.

This allows a direct view of the improvement in multiband parameter estimation accuracy: (1) Among these parameters, the most significant improvement due to multiband is $\mathcal{M}$, with the width of its 90\% CI commonly reaching $10^{-4}$ $M_\odot$, which is consistent with previous results~\citep{Toubiana:2022vpp,Buscicchio:2024asl}. The number of events in this precision interval can be seen from the height of the hist, and some golden events can even reach the order of $10^{-6}$ $M_\odot$. Compared to the ET+2CE network, the addition of LISA can generally improve the accuracy of the chirp mass by 2-3 orders of magnitude. Interestingly, even if LISA observes for only 1 year, the multiband accuracy on chirp mass is still better than the measurement of the ET+2CE network. It was previously generally believed that for sBBH, LISA needs a long observation time to produce good scientific results \citep{Seoane:2021kkk}, but if we use our new method, we can greatly shorten the required time. It is worth noting that in the low measurement accuracy interval of $10^{-3}$ to $10^{-1}$ $M_\odot$, both the multiband and 3G results show a bimodal structure, which we will discuss later; (2) The mass ratio $q$ will also be improved due to LISA, with the common 90\% CI width reaching about 0.03, which is half compared to the 3G results; (3) The effective spin $\chi_\mathrm{eff}$ and asymmetric spin $\chi_\mathrm{a}$ will also be improved due to the information from LISA, where the 90\% CI width of the effective spin $\chi_\mathrm{eff}$ can commonly reach $10^{-3}$ in the multiband case, and even $10^{-4}$ for golden events. In contrast, asymmetric spin $\chi_\mathrm{a}$ is a more difficult quantity to measure, with only a slight improvement in multiband. Similarly to chirp mass, the distribution of asymmetric spin also exhibits a bimodal state, and after verification, its bimodality comes from the bimodality in the chirp mass distribution. This will be discussed in detail later; (4) For extrinsic parameters (right ascension $\alpha$, declination $\delta$, coalesence time $t_c$, luminosity distance $D_L$, inclination angle $\iota$), the improvement from multiband is limited and not of the order of magnitude.

\begin{figure*}
    \centering
    \vspace*{-0.5cm}
    \includegraphics[width=1.0\textwidth]{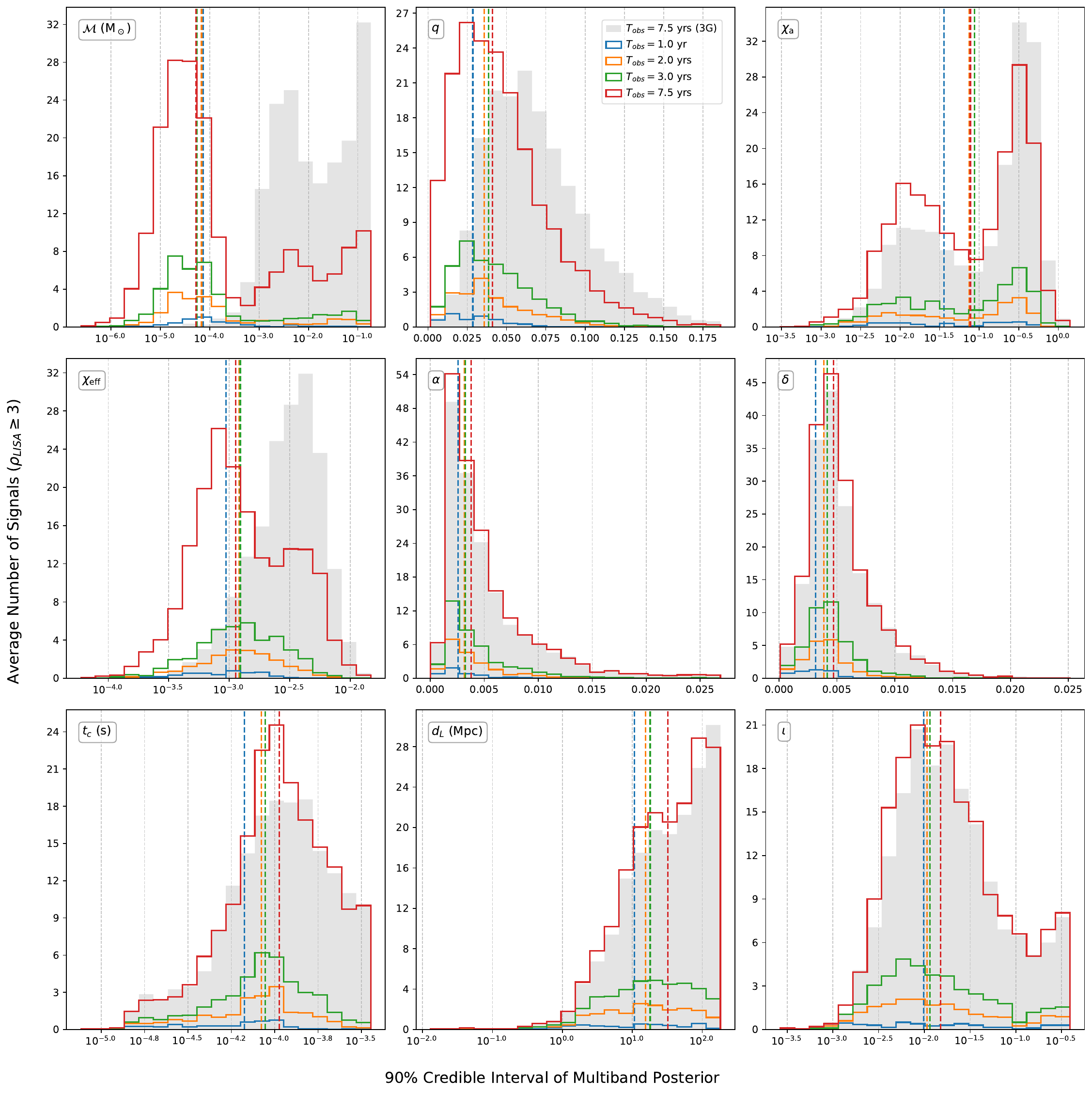}
    \caption{This figure shows the population-scale Bayesian multiband parameter estimation results using our coherent multiband method. Each subplot corresponds to the measurement precision of a parameter. The horizontal axis is the width of the 90\% CI of the parameter's posterior, and the vertical axis is the average number of events within a given measurement precision interval (averaged from 20 random seed datasets). The colored histograms correspond to different LISA observation durations $T_{obs}$ (vertical dashed lines show their median values respectively), and the gray histogram corresponds to the 3G measurement precision for those multiband events. This allows for a direct comparison of whether the addition of LISA improves parameter estimation accuracy. The explanation of bimodal structure in $\mathcal{M}$ and $\chi_{\mathrm{a}}$ can be found in the main text.}
    \label{fig:multiband_pop_ci_width}
\end{figure*}

\begin{figure*}
    \centering
    \vspace*{-0.5cm}
    \includegraphics[width=1.0\textwidth]{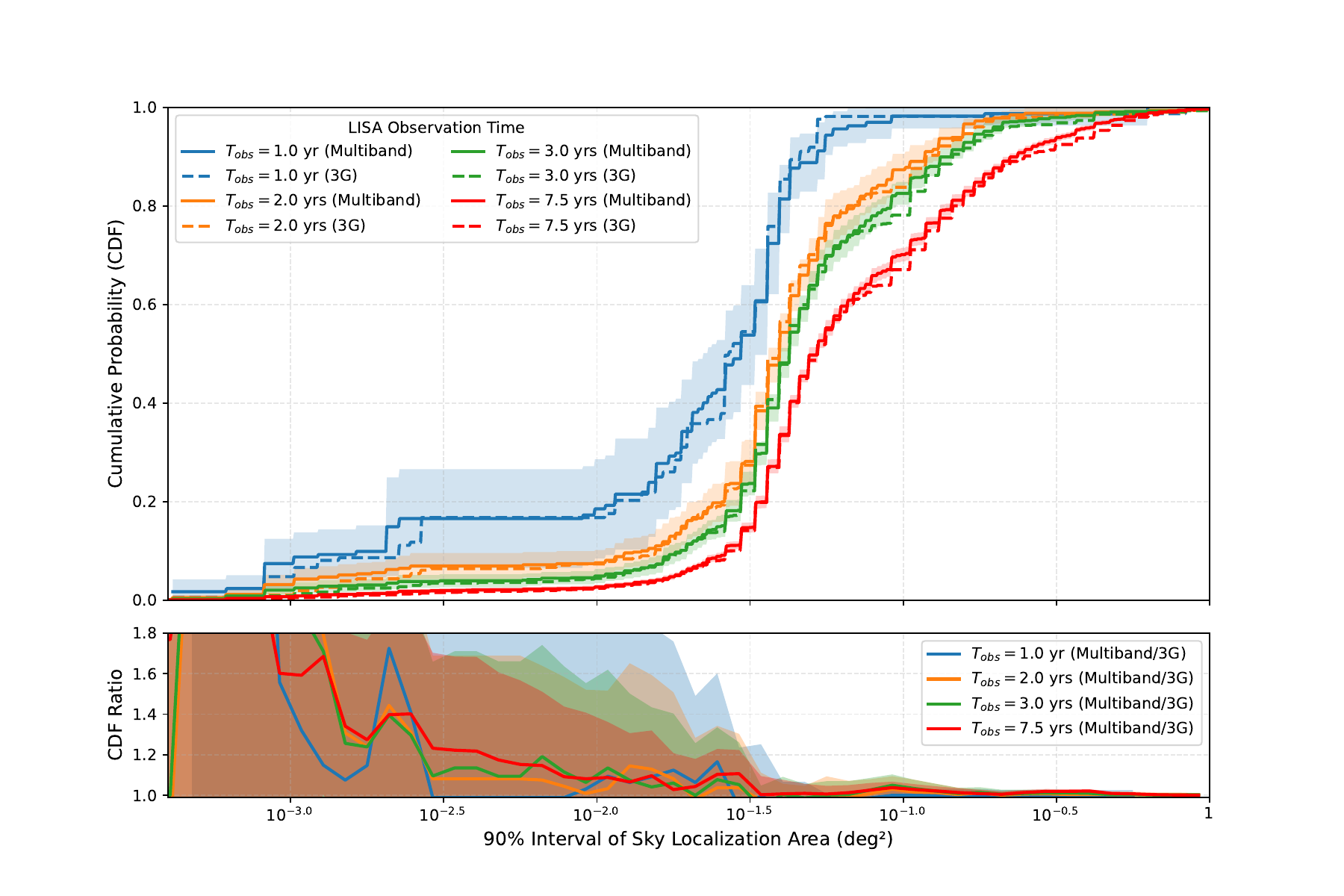}
    \caption{This figure shows the localization capabilities of multiband and 3G for gravitational wave signals from sBBHs. The top panel shows the cumulative distribution of the area of the 90\% credible interval (CI) of the localization posterior for different LISA observation durations. The solid line represents multiband (the shaded area is its 95\% confidence interval), and the dashed line represents the corresponding 3G-only scenario. The longer the observation time, the more the CDF shifts to the right, because the parameter estimation results of much more low-SNR signals are mixed in (worse localization). However, for the same gravitational wave signal, the localization will become better as the LISA observation time increases. The bottom panel shows the ratio of the CDF of multiband to 3G. A value greater than 1 indicates that LISA will help with the localization of sBBHs, although the improvement is marginal.}
    \label{fig:sky_cdf_comparision}
\end{figure*}

Here, we specifically analyze localization in detail because previous literature reported that multiband can make the localization of sBBHs more precise \citep{Klein:2022rbf}, which initially seems inconsistent with our results. Therefore, we calculate the 90\% CI localization area in the celestial sphere based on the posterior samples of the right ascension $\alpha$ and declination $\delta$, as shown in Fig.~\ref{fig:sky_cdf_comparision}. The upper figure shows the cumulative probability density (CDF) distribution of the 90\% localization area. The solid line is the result of multiband (the shaded area is its 95\% confidence interval), and the dashed line is the localization area of the corresponding multiband event only using 3G detectors. It can be seen from the figure that as $T_{obs}$ increases from 1 to 7.5 years, the CDF as a whole moves towards the direction of a larger localization area. Events with a localization area less than $10^{-1.5} \sim 0.0316$ square degrees account for about 60\% of the $T_{obs}=1$ yr signals, about 30\% of the $T_{obs}=2$ yrs signals, about 20\% of the $T_{obs}=3$ yrs signals, and about 15\% of the $T_{obs}=7.5$ yrs signals. The reason why the location of the GW signal appears to be overall better with smaller $T_{obs}$ is due to the selection effect of LISA, which selects more significant signals in such a short time, although the total number of events is small. As the observation time of LISA increases, although the localization accuracy of the same GW signal will be improved to some extent under longer $T_{obs}$, the addition of more and more low-SNR signals makes the overall distribution of localization accuracy move towards a larger value. It can be seen that for the same $T_{obs}$, the CDF of the multiband localization area is almost always in the upper left of the CDF of the 3G localization area, which means that the addition of LISA signals can indeed improve the localization of GW signals to a certain extent. As $T_{obs}$ increases, the difference between the two CDFs in some cases is already greater than the statistical error of the CDF itself. The lower panel in Fig.~\ref{fig:sky_cdf_comparision} shows the ratio of the multiand and 3G localization area CDFs for each $T_{obs}$, which is always greater than 1, and the shaded area is the 95\% error range of the corresponding ratio. It can be seen that multiband can bring improvement to the localization of sBBHs, but it is only marginal.

\begin{figure}[h]
	\includegraphics[width=0.49\textwidth]{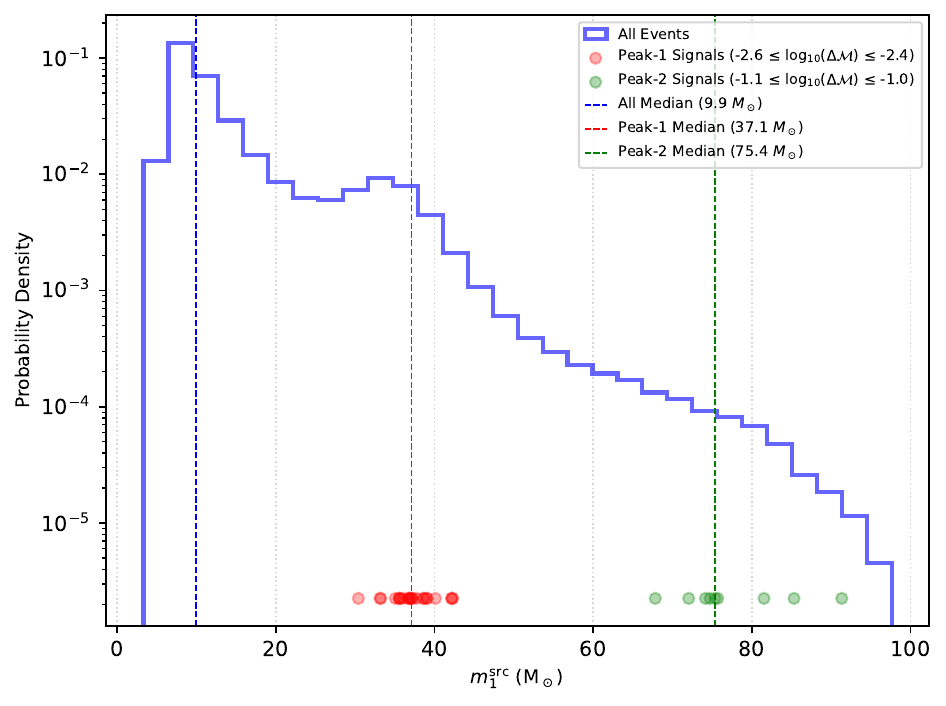}
    \caption{This figure illustrates the population origin of the double peak on the right side of the chirp mass measurement accuracy distribution in Fig.~\ref{fig:multiband_pop_ci_width}. The blue distribution is the \textit{power-law+peak} mass distribution, from which we sampled $m_{1}$ during the signal simulation stage. Since we know the true parameters of all signals, we can trace the origin of those gravitational wave signals on the double peak, and the results show that they come from the \textit{peak} of the \textit{power-law+peak} mass distribution and the heavy tail of the \textit{power-law}, respectively.}
    \label{fig:double_peak_distribution}
\end{figure}

\begin{figure*}
    \centering
    \vspace*{-0.5cm}
    \includegraphics[width=1.0\textwidth]{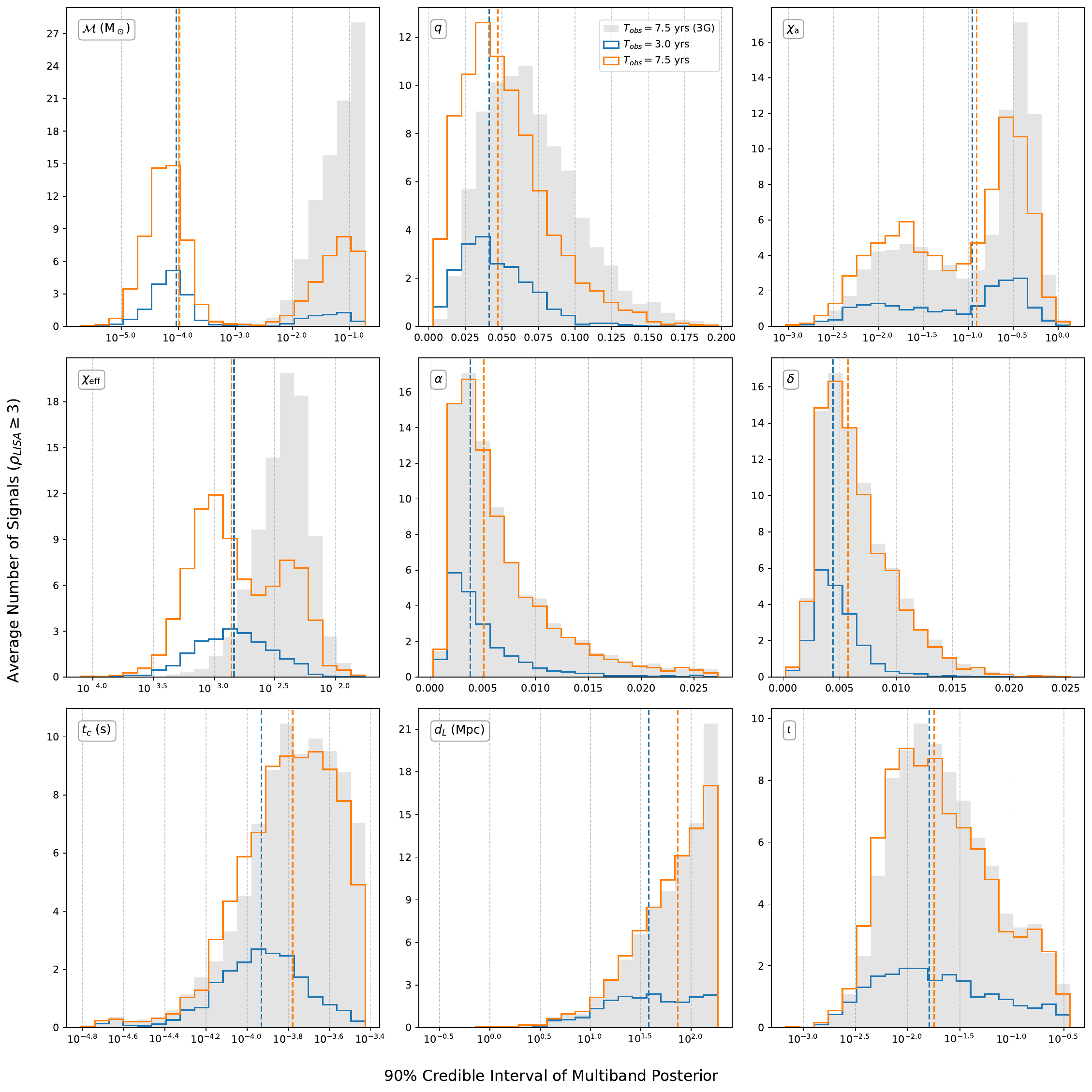}
    \caption{Similar to Fig.~\ref{fig:multiband_pop_ci_width}, but only for IMBHB with source frame total mass $M^{\rm{src}}$ higher than 100$\mathrm{M}_\odot$. Because IMBHB event rate is low in \texttt{GWTC-3} population model, we simulate 50 datasets with different random seeds for averaging to get a statistically meaningful results. In the chirp mass subplot, this result confirms that the right peak ($-2 \le \log_{{10}} (\Delta \mathcal{{M}})$) in Fig.~\ref{fig:multiband_pop_ci_width} is made by IMBHB events, which is consistent with Fig.~\ref{fig:double_peak_distribution}.}
    \label{fig:multiband_pop_ci_width_heavy}
\end{figure*}

Previously, we noticed a bimodal structure in both the multiband and 3G results in the low measurement precision region of $10^{-3}$ to $10^{-1}$ $M_\odot$ on the chirp mass plot in Fig.~\ref{fig:multiband_pop_ci_width}. Does this analysis capture key characteristics of the population? Since we know the true parameters of all these signals, we can trace GW signals located within these two peaks, and the results are shown in Fig.~\ref{fig:double_peak_distribution}. We found that the left peak in Fig.~\ref{fig:multiband_pop_ci_width} ($-3 \le \log_{{10}} (\Delta \mathcal{{M}}) \le -2$) actually corresponds to the \textit{peak} in the \textit{power-law+peak} mass distribution, while the right peak ($-2 \le \log_{{10}} (\Delta \mathcal{{M}})$) in Fig.~\ref{fig:multiband_pop_ci_width} corresponds to the heavy tail of this mass distribution. Although we did not use hierarchical Bayesian inference \citep{Thrane:2018qnx} to infer population parameters, we can still find a clear population feature from the distribution.

We also checked the multiband improvement on light IMBHB events, we selected multiband events with source-frame total mass $M^{\rm{src}} \ge 100\mathrm{M}_\odot$. Considering IMBHB event rate is low in \texttt{GWTC-3} population model, we simulate 50 datasets (different random seeds) to get a statistically stable results. In the chirp mass figure, this result further confirms that the right peak ($-2 \le \log_{{10}} (\Delta \mathcal{{M}})$) in Fig.~\ref{fig:multiband_pop_ci_width} is caused by IMBHB events, which is consistent with results in Fig.~\ref{fig:double_peak_distribution}.

We also evaluate the LISA SNR of GW190521-like signals, specifically those with injected intrinsic parameters within the 90\% CI of the GW190521 posterior \citep{LIGOScientific:2020iuh}. Even with a 7.5-year LISA observation, their LISA SNR remains only around 5, suggesting that heavier IMBHBs are required for higher SNR in LISA band.

\section{\label{sec:conclusions}Conclusions}

We introduce a novel coherent multiband parameter estimation method for the joint analysis of data from future LISA and next-generation ground-based (3G) detectors like ET and CE. This method offers enhanced sensitivity and robustness over previous approaches, enabling the analysis of sBBH signals with lower SNR in the LISA band. Its efficiency allows, for the first time, population-scale multiband Bayesian parameter estimation using simulations based on the \texttt{GWTC-3} model. This significantly expands the potential for multiband-based tests of General Relativity and population inference, effectively doubling the available multiband event rate compared to prior methods. Even with only one year of LISA observations, our method can extract valuable information from sBBH signals, as demonstrated by improvements in parameter estimation (Fig.~\ref{fig:multiband_asd} and Fig.~\ref{fig:multiband_pop_ci_width}).

A key challenge is the low SNR of sBBH signals in LISA (typically $\le$ 8, often $\le$ 5), as shown in Fig.~\ref{fig:optimal_snr_lisa_xg}, contrasting with their very loud signal in ET+2CE (SNR $\sim 500$). For sBBHs near LISA's detection threshold ($\rho_{th}^{\text{LISA}}=5$), the signal's time-domain amplitude is vastly smaller than LISA's noise (panel (a) in Fig.~\ref{fig:strain_snr_sbbh_lisa}), and even matched filtering struggles to distinguish the signal SNR peak from noise SNR peaks (panels (b-c) in Fig.~\ref{fig:strain_snr_sbbh_lisa}).

Our approach (Sec.~\ref{sec:method}), based on phase coherence and extrinsic parameter marginalization, surpasses the $\rho_{th}^{\text{LISA}}=5$ limit of previous methods \citep{Toubiana:2022vpp}. Unlike traditional sequential estimation which propagates systematic errors due to misrepresentation of 3G posterior and is overly sensitive to the prior for low-SNR signals \citep{Vitale:2017cfs}, our method uses rotation matrices to transform signals into a common coordinate system (Sec.~\ref{subsec:coordinate}), ensuring amplitude and phase consistency (at the same reference frequency). We marginalize over extrinsic parameters ($t_{c}$, $\alpha$, $\delta$, luminosity distance, orbital phase) using importance sampling and numerical/analytical integration, aided by precise 3G localization (Fig.~\ref{fig:match_tdi_waveform}), simplifying the exploration to intrinsic parameters. This yields more robust multiband posteriors (Fig.~\ref{fig:comparision_kde_coherent}). We only get the minimal information from LISA-only run, which suffers from low SNR and numerous local likelihood maxima (Fig.~\ref{fig:comparision_blind_coherent}, Fig.~\ref{fig:strain_snr_sbbh_lisa}).

In Sec.~\ref{sec:population_runs}, we validated our method at the population scale using LISA+3G multiband datasets simulated with the \texttt{GWTC-3} population model. Key results include the multiband SNR distribution (Fig.~\ref{fig:optimal_snr_lisa_xg}), event rates under varying SNR thresholds and LISA mission duration (Fig.~\ref{fig:event_rate_lisa}), and 3G detector waiting times for multiband events (Fig.~\ref{fig:merge_time_dist}). We also presented novel analyses on the evolution of sBBH SNR with LISA observation time (Fig.~\ref{fig:lisa_snr_evolution}), and its effects on multiband posteriors for both intrinsic (Fig.~\ref{fig:multiband_different_tobs_intrinsic}) and extrinsic parameters (Fig.~\ref{fig:multiband_different_tobs_extrinsic}).

Our coherent method successfully analyzes all multiband signals in simulated datasets with LISA SNR greater than 3 (Fig.~\ref{fig:xg_multiband_snr_3}), a feat impossible with previous techniques. As a result, we can double the number of available multiband events (Fig.~\ref{fig:event_rate_lisa}).

Based on the analysis of parameter estimation accuracy for all qualified multiband events (Fig.~\ref{fig:multiband_pop_ci_width} and Fig.~\ref{fig:multiband_pop_ci_width_heavy}), we found that detector-frame chirp mass, mass ratio, and effective/asymmetrical spin parameters are greatly improved with LISA's inclusion. However, for the detector settings used, multiband localization shows only marginal improvement over ET+2CE (Fig.~\ref{fig:sky_cdf_comparision}). Notably, LISA's high-precision phase information means that even 1 year of its data, when analyzed with our method, can yield multiband results comparable to or better than the golden events of ET+2CE in 7.5 years of observation (Fig.~\ref{fig:multiband_pop_ci_width}). This revises previous understanding (cf. Table 3.8 in \cite{LISA:2024hlh} and Table 4 in \citep{Seoane:2021kkk}). We also observed that events across the population's mass spectrum correspond to different chirp mass accuracies (Fig.~\ref{fig:double_peak_distribution}), a finding that may aid in understanding underlying sBBH population characteristics through multiband observations.

Nevertheless, waveform systematics are a concern due to our method's reliance on phase coherence. While post-Newtonian models may suffice for quasi-circular, non-precessing, vacuum sBBHs in the LISA band \cite{Mangiagli:2018kpu, LISAConsortiumWaveformWorkingGroup:2023arg}, effects like orbital eccentricity or environmental interactions, potentially negligible for current ground-based detectors \cite{LIGOScientific:2019dag, Romero-Shaw:2019itr, Wu:2020zwr, OShea:2021faf, Ramos-Buades:2023yhy}, become significant in LISA's band. The corresponding high-SNR signals expected in next-generation ground-based detectors \cite{Purrer:2019jcp} will necessitate a three-order-of-magnitude improvement in waveform model accuracy (cf. Fig.~2 in \citep{Purrer:2019jcp}).

Future work will investigate these waveform systematic requirements for our coherent method and apply this method for testing General Relativity or population inference. It would also be interesting if the existing multiband results could be improved by using our method.

The code used in this research is public at \url{https://github.com/gwastro/coherent_multiband_pe}. Our code is based on the \texttt{PyCBC} \citep{Usman:2015kfa, Biwer:2018osg, alex_nitz_2024_10473621}, \texttt{BBHx} \citep{Katz:2020hku}, \texttt{Python} \citep{van1995python}, \texttt{NumPy} \citep{Harris:2020xlr}, \texttt{SymPy} \citep{Meurer:2017yhf}, \texttt{SciPy} \citep{Virtanen:2019joe}, \texttt{Matplotlib} \citep{Hunter:2007}, and \texttt{Excalidraw} \citep{excalidraw}.

% Put together a list of topics that you should discuss

\begin{acknowledgments}
The authors would like to acknowledge Bruce Allen for reviewing the paper draft, Barak Zacky for providing an insightful discussion on the multiband false alarm rate, Sylvain Marsat for providing a useful discussion on the rotation matrix, Michael Katz for providing support of \texttt{BBHx} code, Xisco Jimenez Forteza,  Sumit Kumar, Boris Goncharov, Rahul Dhurkunde and Aleyna Akyuz for useful discussion. The authors thank the computing team from AEI Hannover for their significant technical support. C.C. acknowledges support from NSF Grants No.~PHY-2412341 and AST-2407454.
\end{acknowledgments}

\appendix

\section{Transform between the SSB, LISA, and GEO Frames}
\label{sec:appendix_ssb_lisa_geo}
In the appendix, we will describe how to use rotation matrices to unify the different frames used by LISA and ground-based detectors.

\subsection{\label{subsec:ssb_lisa}Transform between the SSB and LISA Frames}

We can use the following matrix to (inversely) transform between the SSB and LISA frames,

\begin{equation} \label{eq:r_ssb_lisa}
\begin{aligned} 
  &\mathbf{R}_{\text{SL}}(\alpha) \\ % 在 R_SL 前放置对 &
  = \ & % 先写 =，然后紧跟对齐标记 &
      \underbrace{
      \begin{bmatrix}
      \cos\alpha & -\sin\alpha & 0 \\
      \sin\alpha & \cos\alpha & 0 \\
      0 & 0 & 1
      \end{bmatrix}
      }_{\mathbf{R}_z(\alpha)} \cdot
      \underbrace{
      \begin{bmatrix}
      \frac{1}{2} & 0 & -\frac{\sqrt{3}}{2} \\
      0 & 1 & 0 \\
      \frac{\sqrt{3}}{2} & 0 & \frac{1}{2}
      \end{bmatrix}
      }_{\mathbf{R}_y(-\frac{\pi}{3})} \cdot
      \underbrace{
      \begin{bmatrix}
      \cos\alpha & \sin\alpha & 0 \\
      -\sin\alpha & \cos\alpha & 0 \\
      0 & 0 & 1
      \end{bmatrix}
      }_{\mathbf{R}_z(-\alpha)}
\end{aligned}
\end{equation}
with

\begin{equation}
\alpha = \Omega_0 \cdot (t^{\text{LISA}} + t_0),
\label{eq:lisa_angle_alpha}
\end{equation}
the rotation matrix $\mathbf{R}_{\text{SL}}(\alpha)$ transforms the basis vectors of the SSB frame into the basis vectors of the LISA frame. $\alpha$ is the angular position of the LISA centroid in the SSB frame at time $t^{\text{LISA}}$. This paper considers the analytical orbit of LISA \citep{Marsat:2020rtl}, where $\Omega_0$ is the angular velocity of LISA's revolution and $t_0$ is the initial time offset, we set a value to let LISA roughly be behind the Earth by 20 degrees. $\mathbf{R}_z(\alpha)$ represents a rotation around the $z$-axis, which lets the $x$-axis of the SSB frame point to the current position of the LISA centroid. $\mathbf{R}_y(-\frac{\pi}{3})$ represents a rotation around the $y$-axis, which makes the $x-y$ plane of the SSB frame coincide with the LISA detector plane. $\mathbf{R}_z(-\alpha)$ represents a rotation around the $z$-axis, which aims to eliminate the influence of LISA's self-rotation. LISA performs a cartwheel motion in the SSB frame, it revolves around the SSB while also rotating in the opposite direction around its own axis, and both rotation periods are 1 year. Therefore, the magnitudes of the angles in $\mathbf{R}_z(\alpha)$ and $\mathbf{R}_z(-\alpha)$ are equal but have opposite signs. Note that this paper only considers an analytical orbit for LISA; those rotation matrices must be updated to a more complex form when considering a realistic dynamical orbit.

We can use $\mathbf{R}_{\text{SL}}(\alpha)$ to transform the unit propagation vector (location) in the SSB frame to the unit propagation vector (location) in the LISA frame,

\begin{equation}
\hat{\mathbf{k}}^{\text{LISA}} = \mathbf{R}_{\text{SL}}^\top \cdot \hat{\mathbf{k}}^{\text{SSB}}
\label{eq:k_ssb_lisa}
\end{equation}
with

\begin{equation}
\hat{\mathbf{k}}^{\text{SSB}} = 
\begin{bmatrix}
-\cos\beta^{\text{SSB}} \cos\lambda^{\text{SSB}} \\
-\cos\beta^{\text{SSB}} \sin\lambda^{\text{SSB}} \\
-\sin\beta^{\text{SSB}}
\end{bmatrix}
\label{eq:k_ssb}
\end{equation}
and similarly

\begin{equation}
\hat{\mathbf{k}}^{\text{LISA}} = 
\begin{bmatrix}
-\cos\beta^{\text{LISA}} \cos\lambda^{\text{LISA}} \\
-\cos\beta^{\text{LISA}} \sin\lambda^{\text{LISA}} \\
-\sin\beta^{\text{LISA}}
\end{bmatrix},
\label{eq:k_lisa}
\end{equation}
where $\mathbf{R}_{\text{SL}}^\top$ is the transpose of $\mathbf{R}_{\text{SL}}(\alpha)$, because the rotation of the coordinate basis vector is exactly opposite to the rotation of the manipulated vector. $(\lambda^{\text{SSB}},\beta^{\text{SSB}})$ and $(\lambda^{\text{LISA}},\beta^{\text{LISA}})$ are the source localization in the SSB and LISA frames, respectively. We can get them using the corresponding unit propagation vector.

The arrival time of the merge signal from the SSB frame to the LISA frame can be solved iteratively using the following formula,

\begin{equation}
t_c^{\text{LISA}} = t_c^{\text{SSB}} + \frac{\hat{\mathbf{k}}^{\text{SSB}} \cdot \hat{\mathbf{p}}^{\text{SSB}}_\text{{L}}}{c}
\label{eq:tc_ssb_lisa}
\end{equation}
with

\begin{equation}
\hat{\mathbf{p}}^{\text{SSB}}_\text{{L}} = 
\begin{bmatrix}
R_{\text{orbit}} \cos\alpha \\
R_{\text{orbit}} \sin\alpha \\
0
\end{bmatrix},
\label{eq:p_lisa}
\end{equation}
where $\hat{\mathbf{p}}^{\text{SSB}}_\text{{L}}$ is the LISA position vector in the SSB frame at time $t_c^{\text{LISA}}$, because $\alpha$ (Eq.~(\ref{eq:lisa_angle_alpha})) is the angular position of the LISA centroid in the SSB frame at time $t_c^{\text{LISA}}$. $R_{\text{orbit}}=1$ AU is the radius of LISA orbit. $c$ is the speed of light in vacuum. Note that when the signal propagates from the SSB frame's origin to the LISA frame's origin, LISA itself is also moving, so we use $t_c^{\text{LISA}}$ to calculate LISA's position vector $\hat{\mathbf{p}}^{\text{SSB}}_\text{{L}}$ in the SSB frame.

According to the definition of the polarization angle in a given frame \citep{Marsat:2020rtl}, we can get the new polarization angle in the LISA frame,

\begin{subequations}
\begin{align}
\vec{\mathbf{u}}^{\text{SSB}} &= \begin{bmatrix} \sin\lambda^{\text{SSB}} \\ -\cos\lambda^{\text{SSB}} \\ 0 \end{bmatrix} \label{eq:u_ssb} \\
\vec{\mathbf{v}}^{\text{SSB}} &= \begin{bmatrix} 
-\sin\beta^{\text{SSB}} \cos\lambda^{\text{SSB}} \\ 
-\sin\beta^{\text{SSB}} \sin\lambda^{\text{SSB}} \\ 
\cos\beta^{\text{SSB}} 
\end{bmatrix} \label{eq:v_ssb} \\
\vec{\mathbf{p}}^{\text{SSB}} &= \cos\psi^{\text{SSB}} \cdot \vec{\mathbf{u}}^{\text{SSB}} + \sin\psi^{\text{SSB}} \cdot \vec{\mathbf{v}}^{\text{SSB}} \label{eq:p_ssb} \\
\vec{\mathbf{u}}^{\text{LISA}} &= \mathbf{R}_{\text{SL}}^\top \cdot \vec{\mathbf{u}}^{\text{SSB}} \label{eq:u_lisa_rot} \\
\vec{\mathbf{v}}^{\text{LISA}} &= \mathbf{R}_{\text{SL}}^\top \cdot \vec{\mathbf{v}}^{\text{SSB}} \label{eq:v_lisa_rot} \\
\vec{\mathbf{p}}^{\text{LISA}} &= \mathbf{R}_{\text{SL}}^\top \cdot \vec{\mathbf{p}}^{\text{SSB}} \label{eq:p_vec_lisa} \\
\psi^{\text{LISA}} &= \arctan2\left( \frac{\vec{\mathbf{p}}^{\text{LISA}} \cdot \vec{\mathbf{v}}^{\text{LISA}}}{\vec{\mathbf{p}}^{\text{LISA}} \cdot \vec{\mathbf{u}}^{\text{LISA}}} \right) \mod 2\pi \label{eq:psi_ssb_lisa}
\end{align}
\end{subequations}
where $\vec{\mathbf{u}}$ and $\vec{\mathbf{v}}$ are reference polarization vectors, $\vec{\mathbf{p}}$ is one of the polarization basis vectors in the given frame, which is different from those position vectors.

\subsection{\label{subsec:ssb_geo}Transform between SSB and GEO Frame}

Similarly, we can do the transform from the SSB frame to the GEO frame as follows,

\begin{equation}
\mathbf{R}_{\text{SG}}(\epsilon) = 
\underbrace{
\begin{bmatrix}
\cos\epsilon & 0 & \sin\epsilon \\
0 & 1 & 0 \\
-\sin\epsilon & 0 & \cos\epsilon
\end{bmatrix}
}_{\mathbf{R}_x(\epsilon)},
\label{eq:r_ssb_geo}
\end{equation}

\begin{equation}
\hat{\mathbf{k}}^{\text{GEO}} = \mathbf{R}_{\text{SG}}^\top \cdot \hat{\mathbf{k}}^{\text{SSB}},
\label{eq:k_ssb_geo}
\end{equation}

\begin{equation}
t_c^{\text{GEO}} = t_c^{\text{SSB}} + \frac{\hat{\mathbf{k}}^{\text{SSB}} \cdot \hat{\mathbf{p}}^{\text{SSB}}_{\text{E}}}{c},
\label{eq:tc_ssb_geo}
\end{equation}

\begin{equation}
\psi^{\text{GEO}} = \arctan2\left( \frac{\vec{\mathbf{p}}^{\text{GEO}} \cdot \vec{\mathbf{v}}^{\text{GEO}}}{\vec{\mathbf{p}}^{\text{GEO}} \cdot \vec{\mathbf{u}}^{\text{GEO}}} \right) \mod 2\pi,
\label{eq:psi_ssb_geo}
\end{equation}

\begin{equation}
t_d = t_c^{\text{GEO}} + \frac{\hat{\mathbf{k}}^{\text{GEO}} \cdot \hat{\mathbf{p}}^{\text{GEO}}_{\text{3G}}}{c},
\label{eq:td_geo_xg}
\end{equation}
where $\epsilon \approx 0.409095 \ \text{rad}$ is the obliquity of the ecliptic, $\hat{\mathbf{p}}^{\text{SSB}}_{\text{E}}$ represents the position vector of the Earth in the SSB frame, which is roughly 20 degrees ahead of LISA. Using these equations, we can easily perform the (inverse) transformation between two frames. We can use Eq.~(\ref{eq:td_geo_xg}) to calculate the merger time $t_d$ in Eq.~(\ref{eq:inner_product_dh}) for ground-based detectors. $\hat{\mathbf{p}}^{\text{GEO}}_{\text{3G}}$ is the position vector of the 3G detector in the geocentric frame.

\bibliography{apssamp}% Produces the bibliography via BibTeX.

\end{document}